\documentstyle[11pt,aaspp4,psfig]{article}
\newcommand{\etal}{{\it et al.\ }}
\newcommand{\grados}{^\circ}
\lefthead{Aparicio \etal}
\righthead{DDO 187}

\begin{document}

\title{The Star Formation History and the Morphological Evolution of the
Draco Dwarf Spheroidal Galaxy}
%\footnote{Based on observations made with the
%2.5 m Isaac Newton Telescope operated on the island of La Palma by the ING in
%the Spanish Observatorio del Roque de Los Muchachos of the Instituto de
%Astrof\'\i sica de Canarias.}}

\author{Antonio Aparicio} 
\affil{Instituto de Astrof\'\i sica de Canarias,
E38205-La Laguna, Tenerife, Canary Islands, Spain}
\affil{Departamento de Astrof\'\i sica, Universidad de La Laguna,
E38200 - La Laguna, Tenerife, Canary Islands, Spain}

\author{Ricardo Carrera} 
\affil{Instituto de Astrof\'\i sica de Canarias,
E38205-La Laguna, Tenerife, Canary Islands, Spain}

\author{David Mart\'\i nez-Delgado} 
\affil{Instituto de Astrof\'\i sica de Canarias,
E38205-La Laguna, Tenerife, Canary Islands, Spain}

\begin{abstract}

The photometric and morphological properties, as well as the star
formation history, of the Draco dwarf spheroidal galaxy are analyzed on
the basis of wide field, CCD photometry of the resolved stars covering about
$1{\grados}^2$.

Draco is at a distance $d=80\pm 7$ kpc and has a metallicity, [Fe/H], of
$-1.8\pm 0.2$. No metallicity gradient is detected. The star surface density
distribution can be fitted by a single exponential law of scale-length
$\alpha=5\farcm 0\pm 0\farcm 1$. The central surface magnitude is
$\mu_{''V''}=24.4\pm 0.5$ and the core radius $r_{\rm c}= 7\farcm 5\pm
0\farcm 3$ (equivalent to $r_{\rm c}=175\pm 7$ pc). Within errors, the same
scale-lengths are found for the density profiles along the semi-major axis
and the semi-minor axis (re-scaled to semi-major axis units, using the
ellipticity of the galaxy) of Draco. There are hence no
evidences of a tidal-tail associated to Draco. The tidal radius of the
galaxy is found to be $r_t\simeq 42'$ ($\simeq 1$ kpc).

The possibility that the large mass-to-light relation in Draco could be
accounted for by a convenient spatial orientation is tested. An upper limit
to Draco's size along the line of sight is $\sim 14$ kpc. This is too small
to account for the velocity dispersion of Draco if it were due to projection
effects only, and implies that other mechanisms (e.g., dark matter) are
required.

The stellar population of Draco is mainly old. Although some intermediate-age
population is present in Draco, most of the star formation (up to 90\%) took
place before $\sim 10$ Gyr ago. No significant star formation activity is
detected in the last $\sim 2$ Gyr. Two methods ({\it partial model} and {\it
subgiant}) have been used to investigate the star formation history of Drac,
both producing results in good qualitative agreement.

No difference is found between the scale lengths of the distributions of old
($\ga 9$ Gyr) and young ($\sim$ 2--3 Gyr) stars, indicating either that both
populations were formed under the same kinematic conditions, or that any
initial difference was afterwards erased.

\end{abstract}


\section{Introduction}

In contrast with the early, simple idea that dwarf spheroidal (dSph) galaxies
were systems essentially consisting of old, globular cluster-like, Population
II stars (Baade 1963), they have been revealed as objects having undertaken
complex star formation processes during their lives (Mighell \& Rich 1996;
Hurley-Keller, Mateo, \& Nemec 1998; Stetson, Hesser, \& Smecker-Hane 1998;
Mart\'\i nez-Delgado, Gallart, \& Aparicio 1999; Gallart \etal
1999b). Indications that dSphs could contain some intermediate-age or young
stars were early found by Baade \& Swope (1961). Since then, other proofs
have been accumulated, such as the presence of carbon stars (Aaronson \&
Mould 1980; Mould \etal 1982; Frogel \etal 1982; Azzopardi, Lequeux, \&
Westerlund 1986) and bright AGB stars (Elston \& Silva 1992; Freedman 1992;
Lee, Freedman, \& Madore 1993; Davidge 1994; but see also Mart\'\i
nez-Delgado \& Aparicio 1997). Stellar evolution theory became increasingly
important in the interpretation of the color--magnitude diagrams (CMDs) when
these reached the turn-offs of intermediate-age and old stars (see Mould \&
Aaronson 1983 and Mighell 1997 for the Carina dSph and Aparicio 1998 for a
review). Comparison with theoretical isochrones allowed the determination of
the range of ages present in the galaxies. DSph galaxies were studied with
these techniques and the idea that all them, in one way or another, have a
composite stellar population and that no two dSphs could be identified to
have similar star formation histories (SFHs) became popular. In later years,
the most powerful way to obtain the SFH, namely the analysis of the CMD with
synthetic CMDs computed from a stellar evolution library (Aparicio 2001), has
confirmed the wide variety of scenarios and has revealed complex, frequently
bursting SFHs in dSphs (Hurley-Keller \etal 1998; Mighell 1997; Mart\'\i
nez-Delgado \etal 1999; Gallart \etal 1999b; Hern\'andez, Gilmore, \&
Valls-Gabaud 2000; Aparicio 2001).

Structurally, dSphs are usually well fitted by King models, although
exponential fittings are also adequate (Irwin \& Hatzidimitriou 1995). They
show relatively small scale lengths and have low central surface brightness
(Gallagher \& Wise 1994). They are the least-massive galaxies known, but
still, their velocity dispersions imply mass-to-luminosity relations,
$\Upsilon$, as large as $100\ \Upsilon_\odot$. This is usually explained as
dSphs containing large amounts of dark matter, but whether or not they are
really bound systems (Gallagher \& Wise 1994) and whether or not the large
values of $\Upsilon$ are due to peculiar orientation in space (Kroupa 1997;
Klessen \& Kroupa 1998) is still subject of discussion. What is clear is that
at least some dSphs, such as Sagittarius (Mateo, Olszewski, \& Morrison 1998;
Mart\'\i nez-Delgado \etal 2001a) or Ursa Minor (Mart\'\i nez-Delgado \etal
2001b) show evidences of undergoing a process of tidal disruption by the
Milky Way. This would eventually be related to the galaxy formation predicted
in cold dark matter scenarios, in which dwarf galaxies would be the first to
form and would subsequently be merged into larger systems (White \& Rees
1978).

Several efforts have been devoted to the study of Draco. It was discovered by
Wilson (1955) and was the subject of a careful photographic analysis leading
to its first CMD by Baade \& Swope (1961). They found several RR Lyrae
variables and noticed that the horizontal branch (HB) was predominantly
populated in the red part but also had a blue component. They were also the
first to find anomalous Cepheids, later on identified in other dSph
galaxies. Stetson (1979a) published new photographic photometry and estimated
the distance modulus to be $(m-M)_0=19.4$ from the magnitude of the
HB. Deeper CMDs were obtained by Stetson, VandenBergh, \& McClure (1985), who
reached the turn-off of the dominant, old population about 3.5 mag below the
HB.  Carney \& Seitzer (1986) found unambiguous evidence of a blue plume of
stars above the main turn-off, as well as possible traces of an 8 Gyr-old
stellar population. Recently, Grillmair \etal (1998) have presented WFPC2
{\it HST\/} photometry that allows the MS luminosity function down to 3 mag
below the oldest turn-off to be determined.

The stellar density distribution of Draco was early studied by Hodge
(1964) and, more recently, by Irwin \& Hatzidimitriou (1995) who report a
tidal radius of $28'$ and by Lake (1990), Pryor \& Kormendi (1990) and Piatek
\etal (2001) who find evidences of a larger extension of the galaxy.

Draco's metallicity has been found to be very low. From the color of the RGB,
Carney \& Seitzer (1986) estimated it to be [Fe/H] $\simeq -2$ with a
dispersion of 0.8 dex. This is in good agreement with spectroscopic results
for several giant stars, which lead to $-3<[{\rm Fe/H}]<-1.5$ (Stetson 1984;
Shetrone, Bolte, \& Stetson 1998); $[{\rm Fe/H}]=-1.9\pm 0.4$ (Lehnert \etal
1992) or $[{\rm Fe/H}]=-2.00\pm 0.21$ (Shetrone, C\^ot\`e, \& Sargent
2001). This studies also clearly establish a high metallicity dispersion of
the order of 1 dex. 

Since the early work by Baade \& Swope (1961), the SFH of Draco has been the
subject of much discussion. The position and morphology of its RGB would
indicate a very old age for the main population, similar to that of the
globular clusters (Grillmair \etal 1998). But the predominantly red HB seems
to be incompatible with this, at least if age is the second parameter
(although recent results indicate that this assumption might be inadequate:
Rosenberg \etal 1999; Behr \etal 2000). In such a case, Draco should be some
2 Gyr younger than both the globular clusters and the Ursa Minor dSph galaxy
(Stetson \etal 1985; Carney \& Seitzer 1986). On another hand, Carney \&
Seitzer (1986) found some evidence of a younger, $\sim 8$ Gyr-old population
and, if the blue plume is produced by younger MS stars rather than by blue
stragglers, in favor of the hypothesis that the star formation in Draco
should be extended up to a few Gyr ago, even if at a low rate (Carney \&
Seitzer 1986; Grillmair \etal 1998). This would also be compatible with the
presence of the aforementioned anomalous Cepheids, which would belong to a
few Gyr-old population (Norris \& Zinn 1975; Demarque \& Hirshfeld 1975; see
also Gallart \etal 1999a and references therein).

In this paper we present an analysis of the star formation history,
metallicity, and morphological properties of Draco, based on wide-field,
ground-based photometry. We confirm previous results favoring a predominantly
old stellar population and show evidence of a relatively short star formation
burst 2--3 Gyr ago. The paper is organized as follows. In \S 2, the
observations and the data reduction are presented. In \S 3, the CMD of Draco
is presented and the metallicity and distance obtained. In \S 4, the
morphological properties of Draco are discussed, including its overall
structure, surface luminosity distribution and three-dimensional
orientation. In \S 5, the star formation history is computed using two
different methods. Finally, the main results of the paper are summarized in
\S 6.

\section{Observations and data reduction}

\begin{figure}
\centerline{\psfig{figure=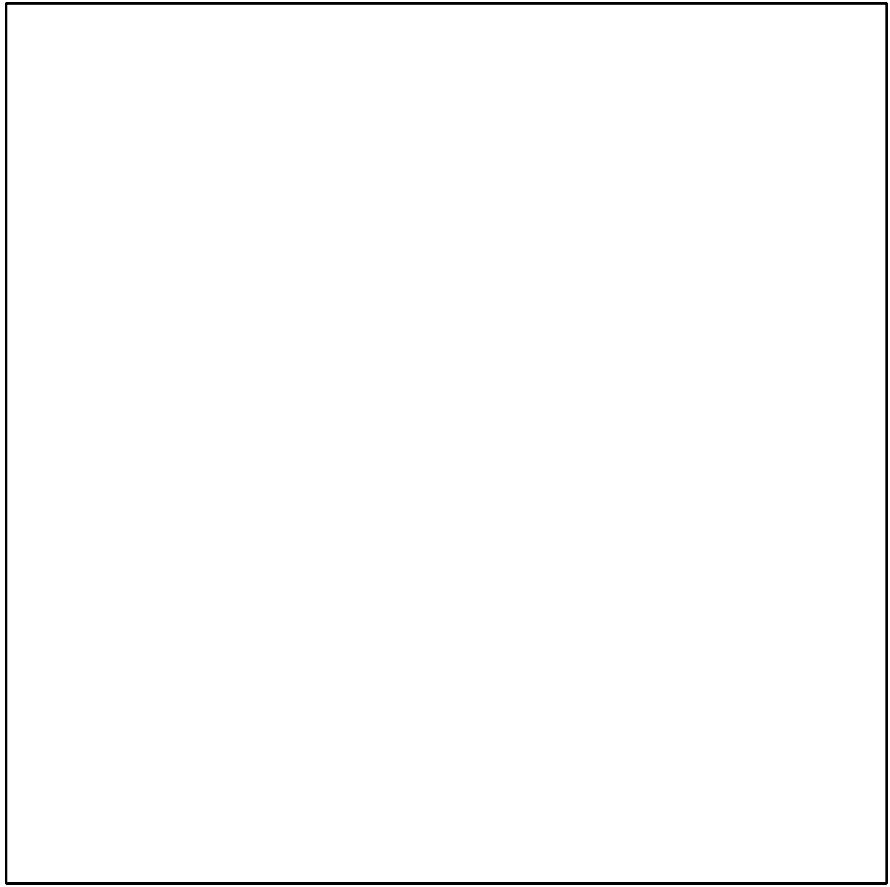,width=16cm}}
\figcaption[ima_1.eps]{Digitized Sky Survey image of Draco. The 
observed fields are shown. North is up, East is left.
\label{ima_1}}
\end{figure}

\begin{figure}
\centerline{\psfig{figure=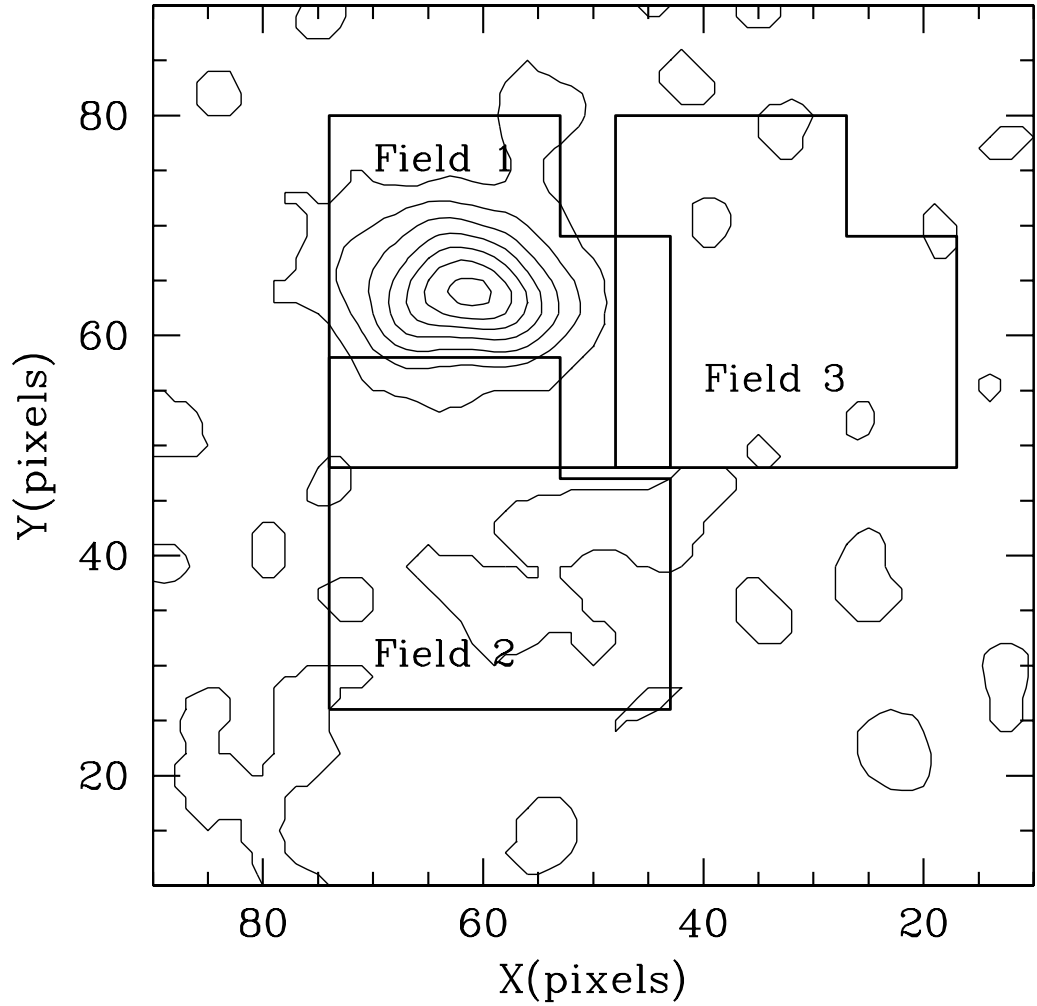,width=16cm}}
\figcaption[ima_2.eps]{Isopleth map of Draco taken from Irwin \&
Hatzidimitriou (1995). Coordinates refer to the scales given by these
authors. The observed fields are shown. North is up, East is left.
\label{ima_2}}
\end{figure}

Images of Draco were obtained in $B$ and $R$ Johnson--Cousins filters during
several observing runs with the INT (2.5 m) at Roque de los Muchachos
Observatory on the island of La Palma (Canary Islands, Spain). Table
\ref{journal} gives the journal of observations. The WFC was used with four
$2048 \times 4096$ EEV chips. The scale is 0.33 $''$/pix which provides a
total field of about $35\times 35$ arcmin$^2$. Several fields were sampled,
as shown in Fig. \ref{ima_1}. Figure \ref{ima_2} displays an isopleth map of
Draco, by Irwin \& Hatzidimitriou (1995), showing also the area covered by
our observations. Total integration times were 1920 s in $B$ and 1600 s in
$R$ for field A (the central field); 3600 s in $B$ and 4200 s in $R$ for field
B, and 2820 s both in $B$ and $R$ for field C.

\placetable{journal}
\placefigure{ima_1}
\placefigure{ima_2}

Bias and flat-field corrections were made with IRAF. Then, DAOPHOT and
ALLFRAME (Stetson 1994) were used to obtain the instrumental photometry of
the resolved stars. Transformation into the standard photometric system
requires several observations of standard star fields in such a way that
standards are measured in all four chips. In practice, we observed three
fields in the four chips and used them to calculate relative photometric
transformations from each chip to the central one (chip 4). Then, a larger
number of standard star fields were measured in chip 4 during the observing
run to calculate both nightly atmospheric extinctions and the general
transformation into the standard Johnson--Cousins system. In total, 280
measurements of 30 standards were made during the observing run of June
1999. Summarizing, the transformations from each chip to chip 4 are:

\begin{equation}
(b_4-b_1)=0.36-0.01(B-R); ~~~~\sigma=0.01,
\end{equation}
\begin{equation}
(r_4-r_1)=0.45+0.04(B-R); ~~~~\sigma=0.02,
\end{equation}
\begin{equation}
(b_4-b_2)=0.409-0.003(B-R); ~~~~\sigma=0.007,
\end{equation}
\begin{equation}
(r_4-r_2)=0.462+0.102(B-R); ~~~~\sigma=0.003,
\end{equation}
\begin{equation}
(b_4-b_3)=0.438-0.024(B-R); ~~~~\sigma=0.008,
\end{equation}
and
\begin{equation}
(r_4-r_3)=0.48+0.06(B-R); ~~~~\sigma=0.03,
\end{equation}

\noindent where subindices refer to chips, lower-case letters stand for
instrumental magnitudes, and capital letters for Johnson--Cousins magnitudes.
The $\sigma$ values are the dispersions of the fits at the centers of mass of
the point distributions; hence they are the minimum internal zero-point
errors. There is a large zero point (bigger than 0.4 mag in most cases)
displacement between chip 4 and the other chips, which implies the lower
sensitivity of the former.

Transformations from chip 4 instrumental magnitudes measured at the top of
the atmosphere into Johnson-Cousins magnitudes are given by:

\begin{equation}
(B-b_4)=24.603+0.036(B-R); ~~~~\sigma=0.006
\end{equation}
and
\begin{equation}
(R-r_4)=24.504+0.040(B-R); ~~~~\sigma=0.010,
\end{equation}

\noindent where, as before, lower-case letters refer to instrumental
magnitudes and capital letters to Johnson--Cousins magnitudes. Finally,
dispersions of the extinctions for each night are about $\sigma=0.01$ for
both filters.

This provides the photometric transformation for the 1999 observing run. For
the remaining runs, transformations were done using as secondary standards
the stars of the Draco fields overlapping with those of the 1999 run. In this
way, the consistency of the photometric scales of the different fields is
secured.

Aperture corrections are necessary to transform the ALLFRAME magnitudes into
large-radius aperture magnitudes. The latter is the system in which the
magnitudes of standard stars have been measured. The transformation has been
performed through aperture photometry on a set of $\sim 200$ isolated bright
stars of field A after subtracting the remaining stars from the
frames. The difference between these and the ALLFRAME magnitudes for the same
stars provides the aperture correction for the corresponding
frame. Dispersions in the aperture corrections are of the order of 0.01.
Putting all the errors together, the total zero-point error of our photometry
can be estimated to be about $\sigma=0.02-0.03$ for both bands, although it
is up to $\sigma=0.04$ for chip 3 in filter $R$.

For each star, ALLFRAME provides with the dispersion in the PSF fitting.  In
general, these dispersions do not reproduce the external errors of the
photometry, which stem mainly from stellar crowding and blending (see
Aparicio, \& Gallart 1995 and Gallart, Aparicio, \& V\'\i lchez 1996), but
they provide an indication of the internal accuracy of the photometry. In
this sense, our average ALLSTAR errors are 0.02, 0.06, and 0.15 for $R=21.1$,
22.8, and 23.9, respectively, and for $B=22.5$, 23.9, and 25.0, respectively.

\begin{figure}
\centerline{\psfig{figure=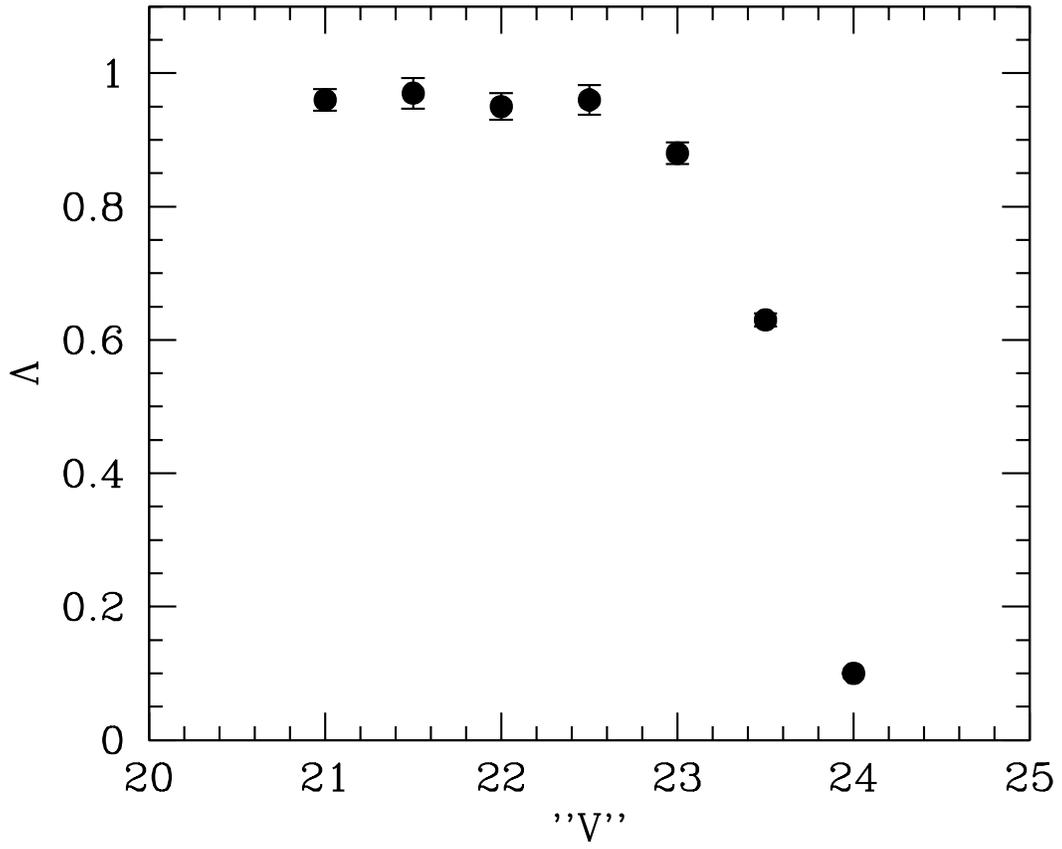,width=16cm}}
\figcaption[crow.eps]{The completeness factor as a function of $''V''$ for
field 1.
\label{crow}}
\end{figure}

Artificial-star tests were performed on field A in the usual way (Stetson \&
Harris 1988; see also Aparicio \& Gallart 1995) to check the observational
effects and estimate completeness factors as a function of magnitude. In
practice, 20~000 artificial stars were added to the $B$ and $R$ images of
Draco. Coordinates of these stars were distributed in a grid covering the
whole image. The step used to build this net was $7''$ both in $x$ and $y$,
avoiding overcrowding effects.  Colors and magnitudes of artificial stars
were obtained from the synthetic CMD (computed as described in Aparicio \&
Gallart 1995) of an old stellar population. In this way, they approximately
mimic the real CMD of Draco, which is important for a realistic observational
effect test. The resulting completeness factors are shown in Figure
\ref{crow}. They start to drop at $''V''\simeq 23.0$ (with $''V''=(B+R)/2$;
see \S 3.1). But about 5\% of the stars are lost at any magnitude brighter
than $''V''\simeq 23.0$. These are mostly stars lying in the regions around
very bright, saturated, field stars.

\placefigure{crow}

\section{The stellar content, metallicity and distance of Draco}

\subsection{The color--magnitude diagram}

Figure \ref{cmd} shows the CMD of Draco. Stars within the inner $30'$
(the semi-major axis) have been plotted (see below). A pseudo-$V$ magnitude,
obtained as $''V''=(B+R)/2$, is used in the vertical axis. This has the
advantage of producing an almost horizontal HB. It shows a well populated,
narrow red giant branch (RGB) with a relatively small slope as well as a
blueward extended HB. These features are the trace of an old, 
low-metallicity stellar population with a small metallicity dispersion. The
main sequence (MS) is visible at $''V''>23$ and, very interestingly, an
extended sequence of blue stars with $''V''<23$, $(B-R)<0.6$ is also
present. This could be a blue-straggler (BS) sequence or the signature of a
younger stellar population.

\placefigure{cmd}

\begin{figure}
\centerline{\psfig{figure=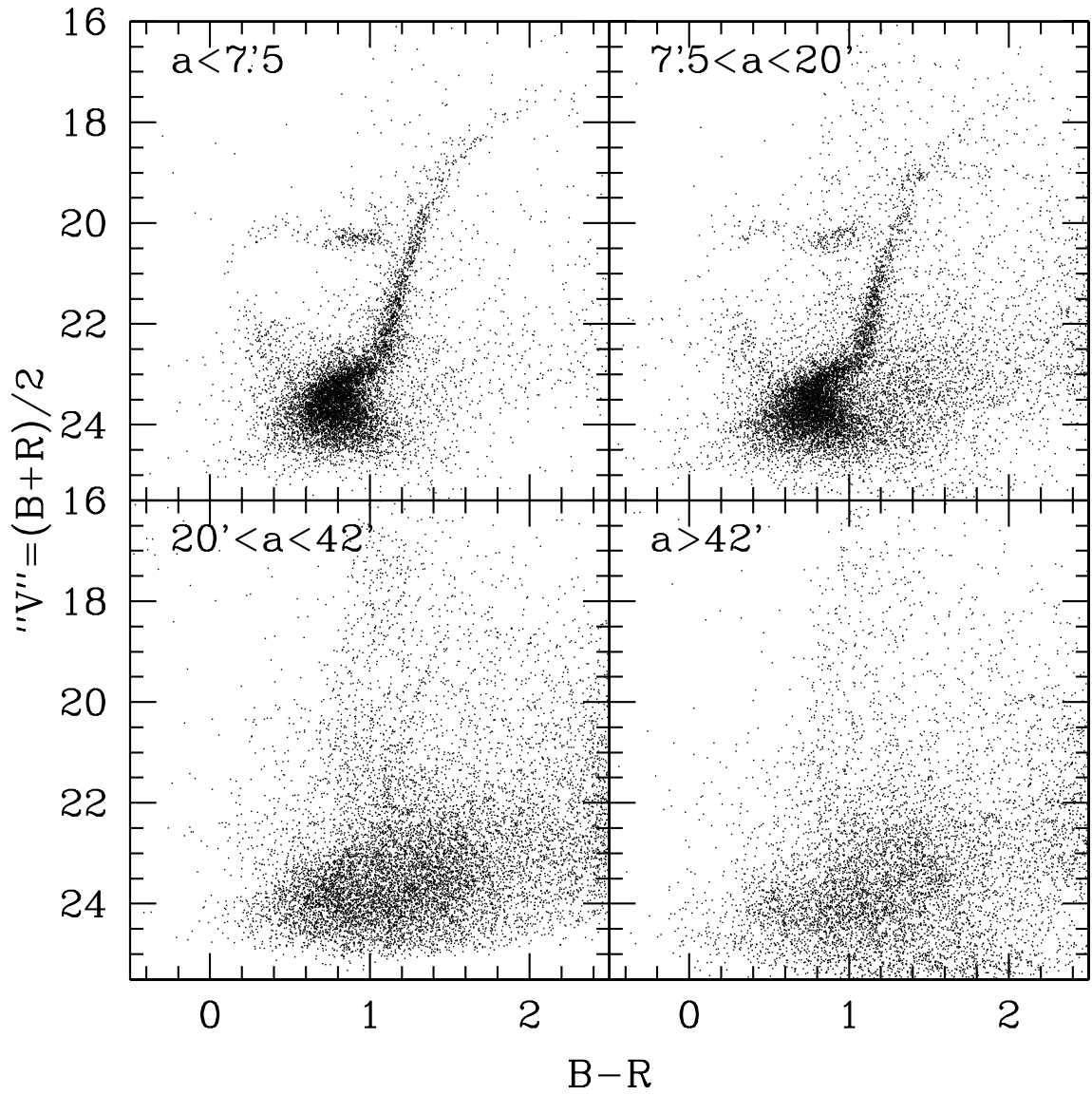,width=16cm}}
\figcaption[cmd_eli.eps]{CMDs of four elliptical annuli centered on the
center of Draco and of semi-major axis as indicated.
\label{cmd_eli}}
\end{figure}

Figure \ref{cmd_eli} shows CMDs for elliptical annuli of increasing
semi-major axis. Note firstly that traces of Draco RGB and MS are still
evident at $20'<a<42'$. Also noticeable is the small group of stars above the
red part of the horizontal branch, visible in the CMD for $a<7/farcm 5$ at
$''V''\simeq 19.7$, $(B-R)=1.0$. This can be a red clump formed by stars
burning He in their nuclei and would be the signature of a
young-to-intermediate population. In fact, Baade \& Swope (1961) and Zinn \&
Searle (1976) found anomalous Cepheids in this region. An explanation for the
origin of these variables is indeed that they are young-to-intermediate age
stars. If so, they should have an MS counterpart, perhaps like the blue
extension discussed above. We will come again to this point in Section 4.

\placefigure{cmd_eli}

These CMDs will be used later on as a reference for discussion
of galactocentric gradients of stellar populations and metallicity.

\subsection{The metallicity}

The metallicity of Draco has been investigated by Carney \& Seitzer (1986),
who obtained [Fe/H] $=-2$ from the color of the RGB. From the spectroscopy of
three giant stars, Shetrone \etal (1998) found that the metallicity should be
in the range $-3.0<[{\rm Fe/H}]<-1.5$. 

The color of the RGB is frequently used to obtain an alternative estimate of
the metallicity. Recently, Gallart et al. (2001) have shown that these values
can be wrong. However we have applied the method to our data and derived an
estimate of the metallicity in the understanding that it will be useful for
comparison with results for other galaxies existing in the literature. 

To do
so, we have used the relations provided by Saviane \etal (2000), based on the
position of the RGB of several globular clusters. However, these relations
are calculated in the $[M_V, (V-I)]$ plane while our photometry is
$BR$. Hence the Saviane {\it et al.}'s relations must be transformed into the
$[M_R, (B-R)]$ plane to be used with our data.

\begin{figure}
\centerline{\psfig{figure=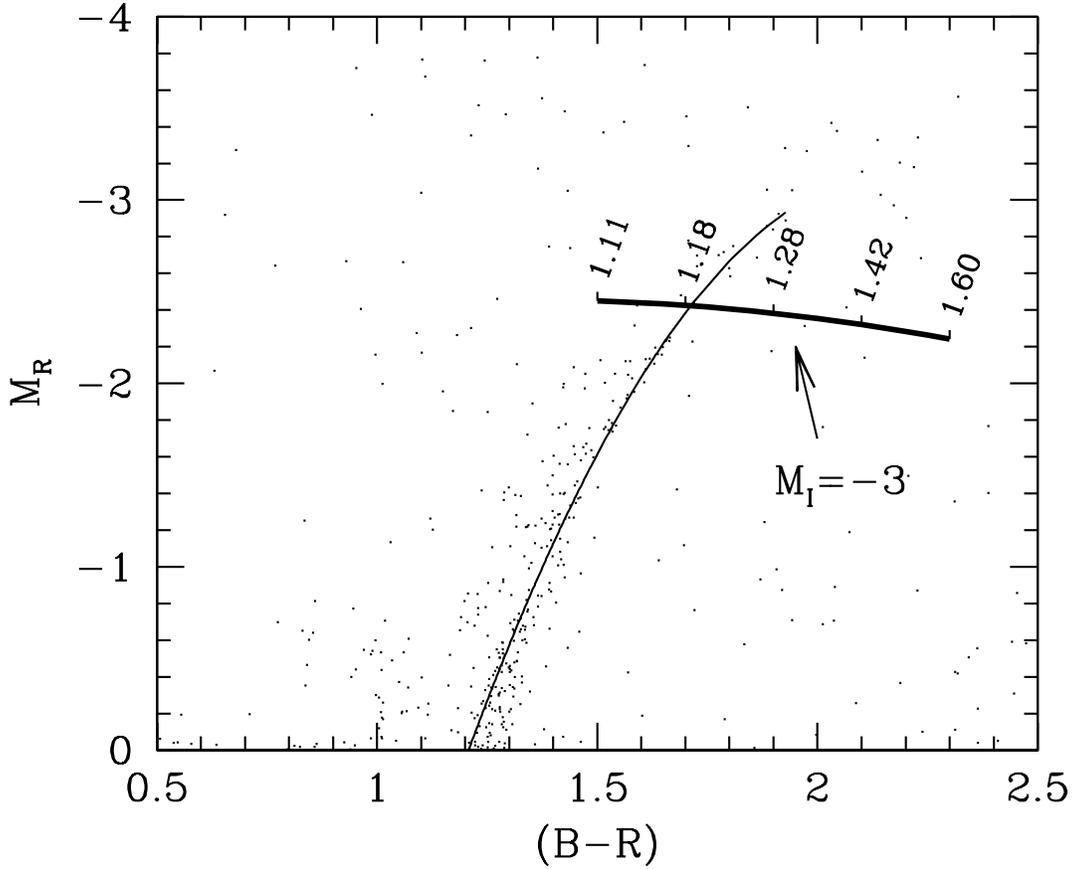,width=16cm}}
\figcaption[cur_met.eps]{The upper part of the Draco RGB. Only stars in the
inner $10'$ have been plotted. The thick track shows the line of constant
absolute $I$ magnitude for $M_I=-3$. The curve has been calculated using a
synthetic CMD. The scale plotted on the line shows the values of
$(V-I)_{I=-3}$ along it. The fiducial RGB of Draco is shown by the thin
line. The point where this line crosses the curve provides $(V-I)_{I=-3}$ for
Draco, from which the metallicity can be estimated.
\label{cur_met}}
\end{figure}

For this we have used the Padua (Bertelli \etal 1994) and Yale (Yi, Demarque,
\& Oemler 1997) isochrones to obtain, for a set of metallicities, the values
of the absolute $M_R$ magnitude and the color indices corresponding to the
RGB at $M_I=-3$. We denote these values by $M_{R_{I=-3}}$, $(V-I)_{I=-3}$ and
$(B-R)_{I=-3}$, respectively. In this way, we obtain the curve in the
[$M_R,(B-R)$] plane shown by a thick line in Figure \ref{cur_met}. Besides
$M_{R_{I=-3}}$ and $(B-R)_{I=-3}$, the parameter changing along this curve is
$(V-I)_{I=-3}$. The intersection of the RGB fiducial line of Draco with the
curve gives its $(V-I)_{I=-3}$, whose value is 1.18. When this is introduced
into the Saviane \etal (2000) relations, the metallicity is obtained to be
[Fe/H] $=-2.0\pm 0.1$ if the Zinn \& West (1984; ZW from here on) metallicity
scale is used, or [Fe/H] $=-1.6\pm 0.1$ if the Carreta \& Gratton (1997; CG
from here on) metallicity scale is used.

\placefigure{cur_met}

Besides this, we have directly obtained an alternative calibration for the
metallicity in the $[M_R, (B-R)]$ plane. For this purpose, we have used the
RGBs of three globular clusters: NGC 1904 (Alcaino \etal 1994), NGC 4590
(Alcaino \etal 1990), and NGC 6723 (Alcaino \etal 1999). A summary of the
metallicity (according to scales for the globular cluster metallicities by ZW
and CG), the distance and the reddening of these clusters is given in Table
\ref{cum}. In accordance with Saviane \etal (2000), the cluster RGBs have
been fitted by a hyperbolic equation of the form $y=a+bx+c/(x-d)$. This fit
has then been used to calculate the color index $(B-R)$ for $M_{R=-2.5}$,
which we will denote $(B-R)_{R=-2.5}$. Finally, the relation between the
metallicity and $(B-R)_{R=-2.5}$ has been determined for both the ZW and CG
scales. For the second case, a linear fit of the form [Fe/H]
$=a(B-R)_{R=-2.5}+b$ is enough, but for the ZW scale, a quadratic fit of the
form [Fe/H] $=a(B-R)_{R=-2.5}^2+b(B-R)_{R=-2.5}+c$ is preferable. The
coefficients of the fits are given in Table \ref{coef}.

\begin{figure}
\centerline{\psfig{figure=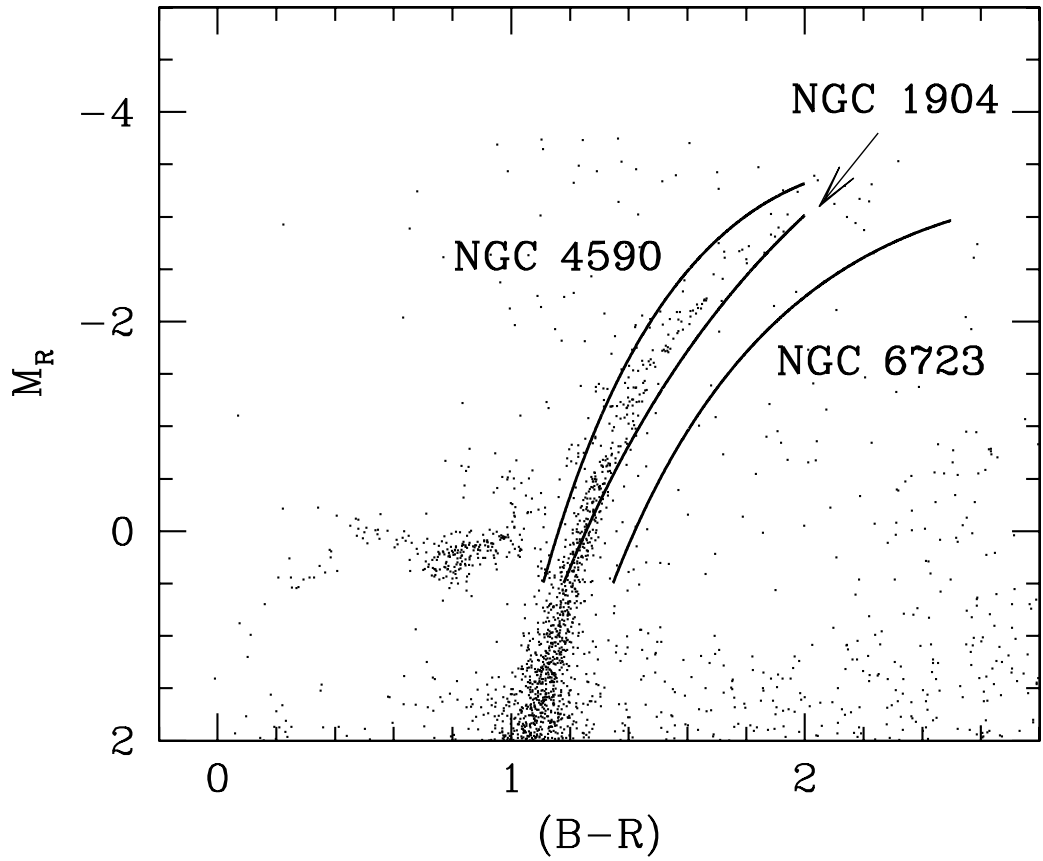,width=16cm}}
\figcaption[superpos.eps]{The fiducial RGBs of the three globular clusters
used to set the metallicity calibration in the $[M_R,(B-R)]$ plane discussed
in the text are shown superimposed on the CMD of Draco (inner $10'$). This
shows that the metallicity of Draco is probably between those of NGC 4590 and
NGC 1904.
\label{superpos}}
\end{figure}

\placetable{cum}
\placetable{coef}
\placefigure{superpos}

The metallicities obtained are [Fe/H] $=-1.9\pm0.1$ if the ZW metallicity
scale is used or [Fe/H] $=-1.6\pm0.1$ if the CG scale is used.

Considering all the values obtained here, we adopt [Fe/H] $=-1.8\pm0.2$ for
the metallicity of Draco. Figure \ref{superpos} shows the RGBs of the three
clusters used to calculate the [Fe/H] $-(B-R)_{R=-2.5}$ relation superimposed
on the CMD of Draco, and demonstrates that, in fact, the metallicity of the
galaxy should be between the NGC 4590 and NGC 1904 values.

\begin{figure}
\centerline{\psfig{figure=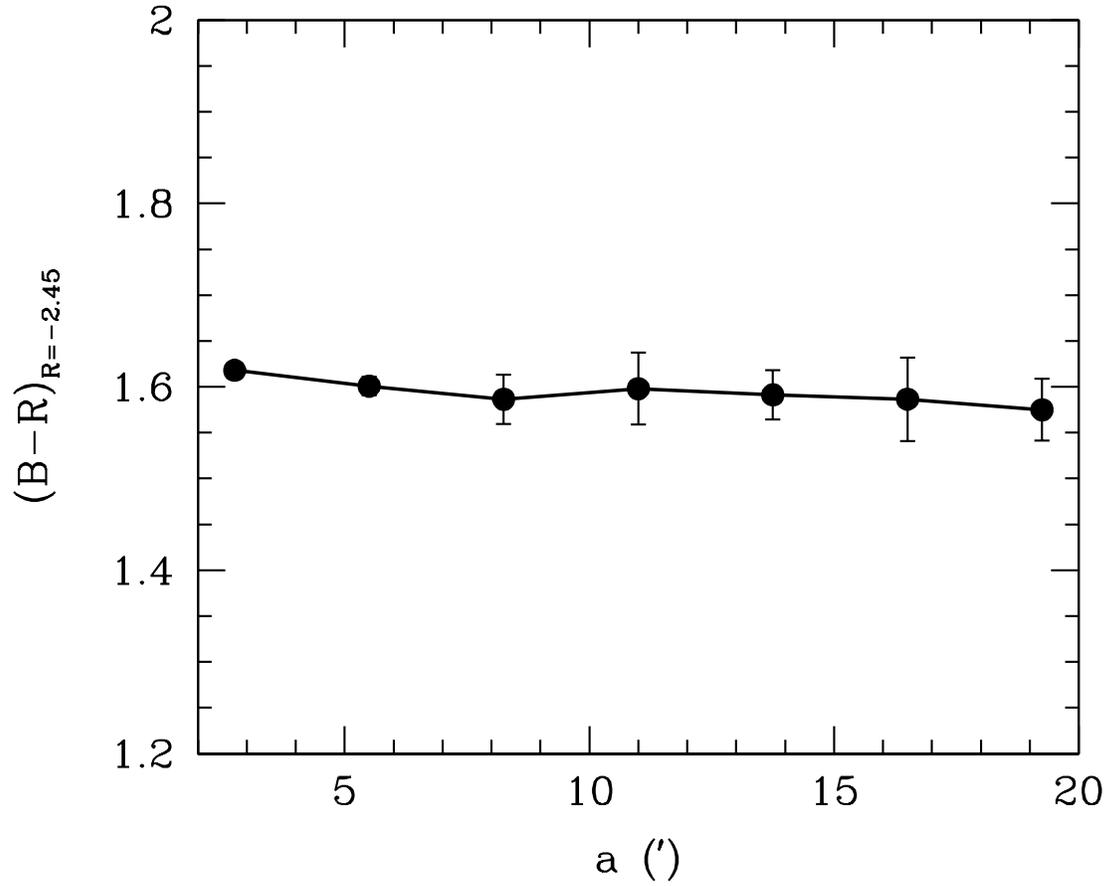,width=16cm}}
\figcaption[met_grad.eps]{$(B-R)_{R=-2.45}$ (the color of the RGB measured at
$M_R=-2.45$) of Draco stars for elliptical annuli of increasing size. If a
metallicity gradient should exist in Draco it would be put into evidence by a
gradient of $(B-R)_{R=-2.45}$.
\label{met_grad}}
\end{figure}

\placefigure{met_grad}

$(B-R)_{R=-2.5}$ and the metallicity obtained therefrom can be used to test
whether galactocentric metallicity gradients exist in Draco. CMDs for
elliptical annuli of increasing semi-major axis, similar to those shown in
Fig. \ref{cmd_eli}, have been used to this purpose. Third-degree polynomial
fits to points in the interval $-1.5\leq R\leq -3.2$ were used to obtain a
fiducial RGB for each CMD. The $(B-R)_{R=-2.5}$ values for each one are
represented in Figure \ref{met_grad}. The error bars show the internal
dispersions of points about the polynomial fits. Inspection of this figure
reveals that no metallicity gradient can be deduced to exist in Draco. 

A metallicity dispersion of about 1 dex in [Fe/H] has been clearly
established in studies referred to the inner $\sim 5'$ of Draco (Stetson
1984; Shetrone, Bolte, \& Stetson 1998; Lehnert \etal 1992; Shetrone,
C\^ot\`e, \& Sargent 2001). Note that this is not in contradiction with the
absence of an overall, large-scale metallicity gradient found here.

\subsection{The distance}

Previous estimates of Draco's distance have been made by Stetson (1979a), who
obtained $d=72\pm 3$ kpc using the position of the red HB, and by Nemec
(1984), who obtained $d=84\pm12$ kpc using RR Lyr variability periods. Using
our data, the distance to Draco can be estimated through the magnitude of the
HB at the RR Lyrae instability strip. From an analytical fit to the blue part
of the HB we have estimated that magnitude to be $''V''=20.2\pm 0.1$. A shift
has to be applied to this value to obtain the actual $V$
magnitude. Considering the $BVR$ magnitudes in the RR Lyrae instability strip
given by the Padua stellar evolution library, that shift is $''V''-V=0.06$,
hence $V=20.14$. Finally, using the relation by Lee \etal (1999) for the
luminosity-metallicity of the HB, $M_V=0.17$[Fe/H]$+0.82$, the distance
becomes $(m-M)_V=19.6\pm 0.2$, where the quoted error includes uncertainties
in the $''V''-V$ transformation, as well as the metallicity error. Reddening
for Draco was investigated by Stetson (1979b), who found $E(B-V)=0.03\pm
0.01$. It can also be calculated from the IR maps of dust by Schlegel,
Finkbeiner, \& Davis (1998), which give $E(B-V)=0.028$, corresponding to an
extinction of $A_V=0.09$. Applying this correction to the distance modulus
gives $(m-M)_0=19.5\pm 0.2$, or $d=80\pm 7$ kpc. We will adopt this value.

\section{The morphological properties of Draco}

\subsection{Density profile}

The radial stellar density distribution of Draco was first obtained from
wide-field photographic photometry by Hodge (1964), who reported a tidal
radius $r_t=26'$, and more recently by Irwin \& Hatzidimitriou (1995), who
derived $r_t=28'$. Lake (1990) and Pryor \& Kormendi (1990) modeled the
galaxy using Hodge's data and found $r_t=55'$ and $r_t=36'$,
respectively. Very recently, Piatek \etal (2001) have carried out an
extensive search of Draco stars beyond the Irwin \& Hatzidimitriou (1995)
tidal radius, finding evidence of the existence of such stars.

Here we provide a new profile, now based on CCD photometry. Three stellar
distributions have been obtained from star counts in elliptical annuli of
increasing semi-major axis. The eccentricity of the elliptical annuli has
been assumed to be $\epsilon=1-b/a=0.29$ and the position angle 82$^\circ$,
after Irwin \& Hatzidimitriou (1995). The center of the galaxy given by these
authors has also been adopted. The three density profiles have been computed
as follows. For the first one, all the resolved stars have been used. For the
other two, only stars in two regions of the CMD where the ratio of Draco to
foreground stars is largest, have been considered: the MS and the RGB. To
sample the MS, stars within the box defined by $0.6<(B-R)<1.0$;
$24.25<''V''<22.75$ have been used. For the RGB, box 14, as defined in Figure
\ref{cmd_box}, has been used.

\begin{figure}
\centerline{\psfig{figure=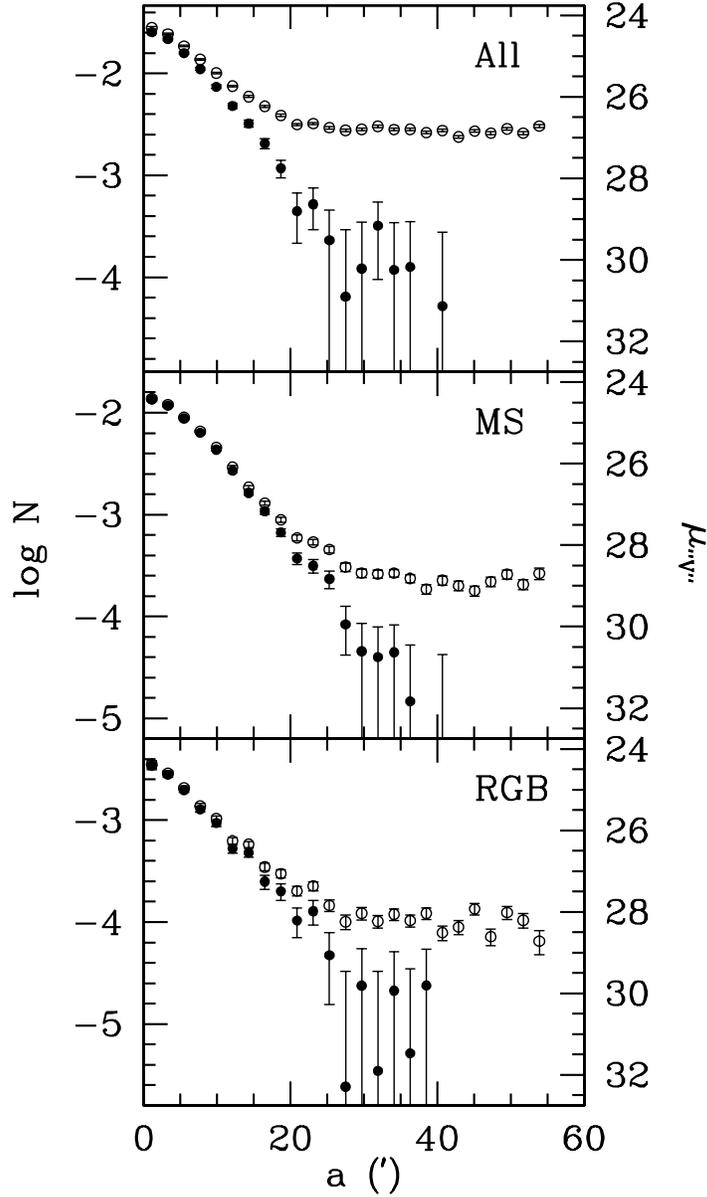,width=16cm}}
\figcaption[densi.eps]{Surface star density and surface magnitude
distributions in Draco. All the resolved stars have been considered to plot
the upper panel figure, while only MS and RGB stars have been respectively
used for the medium and lower panel figures. Stars have been counted in
elliptical annuli of eccentricity $1-a/b=0.29$, position angle $82^\circ$ and
increasing semi-major axis. Horizontal axis gives the semimajor-axis in
arcmin. Left-hand vertical axis gives the logarithm of the star density per
arcsec$^2$. The right-hand vertical axis gives an estimate of the surface
magnitude, calibrated as described in text. Open circles represent the raw
density distribution. Filled circles show the density distribution after
subtraction of the sky foreground contamination, estimated as described in
text.
\label{densi_p}}
\end{figure}

\begin{figure}
\centerline{\psfig{figure=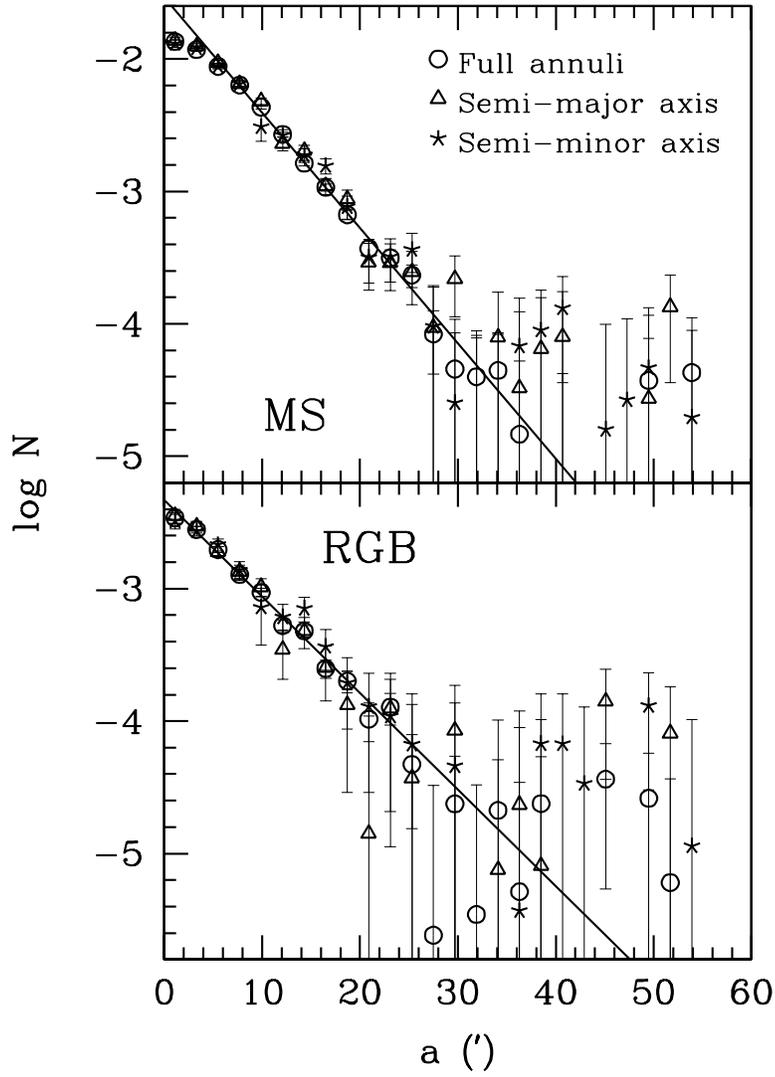,width=16cm}}
\figcaption[densi.eps]{Surface star density of stars along the semi-major and
semi-minor axis of Draco, after correction of foreground contamination. MS
(upper panel) and RGB (lower panel) stars have been considered. The density
distribution obtained for star counts in the full elliptical annuli are also
plotted for comparison. Straight lines show the fits done using full
elliptical annuli distributions.
\label{densi_3}}
\end{figure}

\placefigure{densi_p}
\placefigure{densi_3}

The final star density distributions are shown in Figure \ref{densi_p}. The
corresponding $''V''$ surface magnitude is given in the right-hand vertical
scale, according to the magnitude scale obtained below. In all the cases,
open circles show the raw distributions. The sky contamination by foreground
and background objects must be subtracted from this to obtain the
distribution of Draco stars. Background, extended objects are automatically
removed by DAOPHOT, while background, stellar-shaped objects are
indistinguishable from foreground stars. For this reason, from here on, we
will refer to foreground contamination only, understanding that it include
also the background stellar-shaped objects not rejected by DAOPHOT.

The raw radial surface density distributions (Fig. \ref{densi_p}, open
circles) change slope at $a\simeq20'-25'$, but continue decreasing until at
least $a\simeq 42'$. Let assume that the sky star density has been reached at
this distance. If the average densities for $a>42'$ are subtracted from the
raw radial distributions, the profiles shown by full dots in Figure
\ref{densi_p} are obtained. A single law can account for the density profile
for $a>5'$ in all the cases. On the one hand, the absence of a plateau for
large galactocentric distances indicates that a tidal tail is not present in
Draco. On the other hand, this indicates that the tidal radius (semi-major
axis) of Draco is about $r_t=42'$.

A way to further explore the presence of extra-tidal stars or of a tidal tail
structure is to compare the density profiles in the direction of the major
and minor axes. We have done so for the MS and the RGB stars, for which the
Draco-to-foreground star ratio is greater. The results are shown in Figure
\ref{densi_3}. The distributions obtained from star counting in full ellipses
(the same shown in Fig. \ref{densi_p}, full dots) are compared with those for
stars within two strips along the western semi-major axis and the southern
semi-minor axis. Table \ref{escalas} lists the scale-lengths in each case
(column 2). For the semi-minor axis, these quantities have been scaled into
units along the semi-major axis; i.e., they have been divided by $b/a$, so
that they are directly comparable with the scale-lengths corresponding to the
semi-major axis and full ellipses. No significant differences are found in
either case, which leads to the conclusion that Draco is a very extended,
low density galaxy, but that no evidence of a tidal tail populated by stars
stripped away from the galaxy are found. In any case, more data about the
stellar population in fields around Draco's main body are required to better
understand the galaxy's structural properties.

Finally, the core radius (semi-major axis) is $r_{\rm c}=7\farcm 5\pm 0\farcm
3$. At the distance adopted for Draco, this corresponds to $r_{\rm c}=175\pm
7$ pc. 

\subsection{Surface photometry}

It is difficult to obtain the surface brightness distribution of galaxies
which, like Draco, have a very low stellar density and which are resolved
down to stars of quite faint intrinsic luminosity. An indirect technique is
to obtain the surface density distribution of resolved stars, which is
afterwards transformed into a surface-brightness scale. In our case, this
transformation has been done as follows. Firstly, a synthetic model for a
stellar population as found for Draco in \S 5 has been computed. Secondly,
the number of RGB stars has been counted in this model. To do this, stars in
box 14 (see Figure \ref{cmd_box}) have been considered. Finally, comparison
of this number with the total luminosity in $B$ and $R$ for the synthetic
model provides a scale transformation from RGB star counts to magnitudes. The
accuracy of the final scale of magnitudes cannot be expected to be high. In
particular, it depends on the assumed IMF. But it is probably better than that
obtained through direct measurement of brightness on a big field dominated by
background noise and contamination by bright foreground stars.

Applying this method, we estimate the central surface brightness in $B$, $R$,
and $``V''$ to be $\mu_{B,0}=25.0\pm 0.5$, $\mu_{R,0}=23.9\pm 0.5$ and
$\mu_{``V'',0}=24.4\pm 0.5$, where the quoted errors should account for the
high degree of uncertainty. Irwin \& Hatzidimitriou (1995) obtained
$\mu_{V,0}=25.5\pm 0.5$, which is one magnitude fainter than ours. It is
difficult to say which value is preferable. Irwin \& Hatzidimitriou have
obtained theirs from a King model fit to photographic data. Our CCD
photometry is deeper and is expected to be more accurate, but, as previously
quoted, our estimate could be affected by a wrong choice of the IMF and, in
general, by the fact that we are using a stellar population model. In our
opinion, the relatively large disagreement between both values reveals
nothing but the difficulty of obtaining accurate surface photometry of this
kind of systems.

The integrated magnitudes of Draco can be obtained by integration of the
density profile and the central surface brightness abovely obtained. The
results, together with all the morphological parameters obtained here and
several other integrated quantities are summarized in Table
\ref{densi_par}. In Table \ref{par_com}, morphological parameters are
compared with other authors' results.

\placetable{densi_par}
\placetable{par_com}

\subsection{Spatial distribution of stellar populations}

The spatial distribution of stellar populations of different ages gives
information on the time-scale in which stars mix out in the galaxy and on
kinematic and/or morphological evolution. Computing the detailed models
required to perform an accurate analysis is beyond the scope of this
paper. But valuable information can be obtained from the comparison of the
radial distribution scale lengths of populations of different ages. The latter
can be selected from the boxes defined in Fig. \ref{cmd_box}, since
different boxes sample different ages (see details in \S 5).

From synthetic models computed for the SFH study (see below, \S 5) it can be
seen that boxes 12 and 14 (Fig. \ref{cmd_box}) contain stars with ages in the
interval 9--15 Gyr, while boxes 2 to 6, 10 and 13 contain younger stars and
preferently, stars in the interval 2--3 Gyr (they sample the small recent
starburst; see \S 5). We have calculated the radial distribution scale length
for both groups of stars in the region $5' \leq a\leq 20'$. We have obtained
$\alpha=5\farcm 8\pm 0\farcm 2$ for the older population and $\alpha=6'\pm
1'$ for the younger; {\it i.e.,} there is no significant difference either
between both or with respect to the whole resolved stellar population.
Within the errors, this indicates that stars of different ages where formed
with similar spatial distribution or that their movement in the galaxy has
completely mixed them.

\subsection{On the three-dimensional orientation of Draco}

Dwarf spheroidal galaxies usually present high mass-to-luminosity ratios
($\Upsilon$). Draco, in particular, has $\Upsilon=84\ \Upsilon_\odot$ (Mateo
1998). Dark matter is usually invoked as an explanation for these large
values in a virialized scenario. An alternative explanation has been proposed
by Kroupa (1997) and Klessen \& Kroupa (1998), who suggest that tidal
disruption of these galaxies by the Milky Way potential would well produce
very elongated systems whose main axis, together with the main axis of the
velocity ellipsoid, could eventually be oriented close to the line of
sight. In such a case, the high observed velocity dispersion would not need
of large quantities of dark matter to be accounted for.

\begin{figure}
\centerline{\psfig{figure=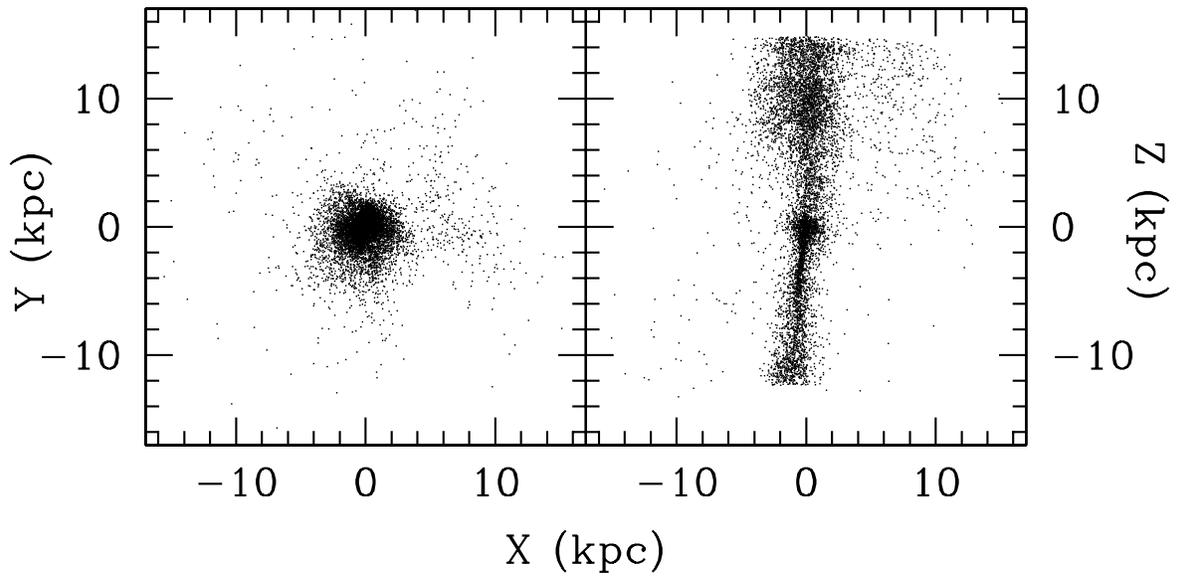,width=16cm}} 
\figcaption[kroupa.eps]{The
spatial distribution of stars according to RS1-4 model of Kroupa (1997),
which reproduces the velocity dispersion of Draco. The $Z$ axis corresponds
to the line of sight. Coordinates are given in kpc and are centered on the
center of Draco.
\label{kroupa}}
\end{figure}

\begin{figure}
\centerline{\psfig{figure=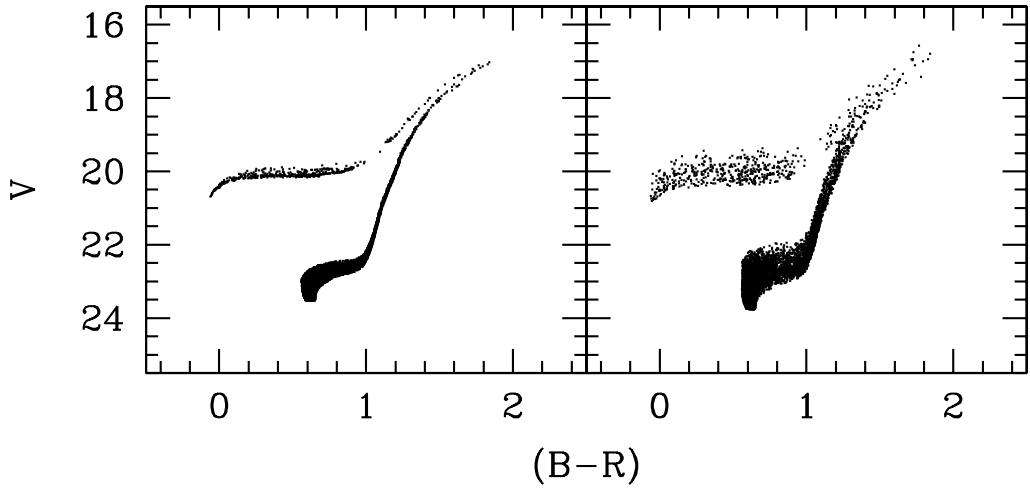,width=16cm}} 
\figcaption[disper.eps]{The
left-hand panel shows a synthetic CMD for an old stellar population, similar
to the main population found in Draco (see \S 4). The right-hand panel shows
the same population including the simulation of the distance dispersion
corresponding the RS1-4 model of Kroupa (1997), as shown in Figure
\protect\ref{kroupa}.
\label{disper}}
\end{figure}

\placefigure{kroupa}
\placefigure{disper}

To check this possibility we have adopted the hypothesis that Draco is
configured according to model RS1-4 of Kroupa (1997), which is the one that
most closely reproduces the $\Upsilon$ of Draco. The spatial distribution of
stars in this model is shown in Figure \ref{kroupa}. We have then simulated
the predicted distance dispersion of Draco stars to us in a synthetic CMD of
a stellar population similar to that found for Draco (see \S4). The CMDs
before and after simulating the distance dispersion are shown in Figure
\ref{disper}. Comparison of the right-hand CMD with the Draco CMD
(Fig. \ref{cmd}) clearly indicates the incompatibility of the RS1-4 model
with the observations.

On another hand, the extension of Draco along the line of sight can be
estimated from the width of the HB. This width is potentially affected by
several factors, such as observational effects, stellar evolution or binarity
of composite stellar populations. From our artificial-star test (\S 2) we
have checked that widening of the HB by observational effects can be
neglected, but considering every other possible effect in detail is
complicated. However, an upper limit estimate can still be set by neglecting
all the effects and assuming that the entire width observed in the red HB
(Fig. \ref{cmd}) is produced by distance dispersion. In such a case, the size
of Draco along the line of sight would be $\sim 14$ kpc. This is quite a
large value compared with Draco's size in the plane of sky but it should be
stressed that it is an upper limit, and that it is still less than 1/3 the
size given by the RS1-4 model. 

\section{The Star Formation History}

The SFH of a galaxy can be derived in detail from a deep CMD through
comparison with synthetic CMDs (Aparicio 2001). Details on the
computation of synthetic CMDs and on the method used here to derive the SFH
can be found in Gallart \etal (1999b), Aparicio \& Gallart (1995), and
references therein. The accuracy and temporal resolution of the SFH depend on
the quality and depth of the observed CMD and on the uncertainties in the
stellar evolutionary library upon which the synthetic CMDs are computed. To
obtain a reasonable resolution for old ages the CMD must sample stellar
evolutionary phases sensitive to these ages, such as the HB or the lower
MS. Moreover, the RC, the RGB, and the HB itself provide  information on
the chemical enrichment law (CEL).

The CEL and the initial mass function (IMF), together with the star formation
rate (SFR), are the functions contained in the more general SFH. A function
containing information on the fraction and mass distribution of binary stars
would also be relevant. For simplicity, we have neglected here the influence
of binary stars and have assumed a fixed IMF: that given by Kroupa, Tout, \&
Gilmore (1993). The CEL can be constrained by the morphology and color of the
RGB. As discussed above (see \S3.2) Draco presents a low average metallicity
and a low metallicity dispersion, [Fe/H] $=-1.8\pm 0.2$. This corresponds to
$0.0002\leq Z\leq 0.0003$. However, the Padua stellar evolutionary models
produce a best fit to Draco RGB for $Z\simeq 0.0005$. This difference is due
to the internal tuning of the stellar evolution models and is not important
for the analysis of the metallicity. However, it is preferable using it to
compute the synthetic CMDs necessary for the SFH study. In practice, we have
assumed a CEL independent of time producing metallicity values for the stars
randomly distributed in the interval $0.0004\leq Z\leq 0.0006$ for all
ages. Provided that the stellar population does not have a large amount of
young stars (younger than some 1 Gyr), as can be deduced from the extension
of the MS of Draco, this CEL produces synthetic CMDs with an RGB always
compatible with that of Draco.

The only remaining function to complete the SFH of Draco is the SFR. Two
different methods have been used to obtain it. The first method is that of
the IAC-Padua group (see Aparicio 1998 for a description and Gallart \etal
1999b for one of the latest applications), which we will refer to as the
{\it partial diagram} or {\it partial model} method. The second method is
that presented by Mighell (1997), which we will refer to as the {\it
subgiant} method. We will discuss in turn the results obtained with each
of these.

\subsection{The partial diagram method}

This method was introduced by Aparicio, Gallart, \& Bertelli
 (1997). To apply it, a few
synthetic CMDs (partial models) are computed, each for a stellar
population with ages in a narrow interval (typically, a few partial
models are used each embracing an age interval ranging from several $10^8$
yr to a few Gyr to cover the full $\sim 15$ Gyr of galaxy ages). The Padua
stellar evolution library (Bertelli \etal 1994) is used for this purpose. The
basic idea is that any SFR can be simulated as a linear combination of
partial models.

The practical way to derive the SFR is as follows. A number of boxes are
defined in the CMDs (observed and partial models) and the number of
stars are counted inside them. The boxes are defined in such a way they
sample stellar evolutionary phases providing information about different
ages. Let us call $N_j^o$ the number of stars in box $j$ in the observed CMD and
$N_{ji}^m$ the number of stars in box $j$ of partial model $i$ (the
partial model covering the $i$th age interval). Using this, the
distribution of stars in boxes of an arbitrary SFR can be obtained from a
linear combination of the $N_{ji}^m$, by
\begin{equation}
N_j^m=k\sum_i\alpha_iN_{ji}^m.
\end{equation}
\noindent The corresponding SFR can be written as
\begin{equation}
\psi(t)=k\sum_i\alpha_i\psi_p\Delta_i(t),
\end{equation}
\noindent where $\alpha_i$ are the linear combination coefficients, $k$ is a
scaling constant, and $\Delta_i(t)=1$ if $t$ is inside the interval
corresponding to partial model $i$, and $\Delta_i(t)=0$ otherwise. Finally,
the $\psi(t)$ having the best compatibility with the data can be obtained by
a least-squares fitting of $N_j^m$ to $N_j^o$, the $\alpha_i$ coefficients
being the free parameters. 

The finite time interval of partial models introduces a limitation on the
time resolution of the SFH. But it must be noted that the final limitation is
indeed imposed by the accuracy of observations and the precision of stellar
evolutionary models and also that increasing the time resolution of the
computed solution is only a matter of computer time.

\begin{figure}
\centerline{\psfig{figure=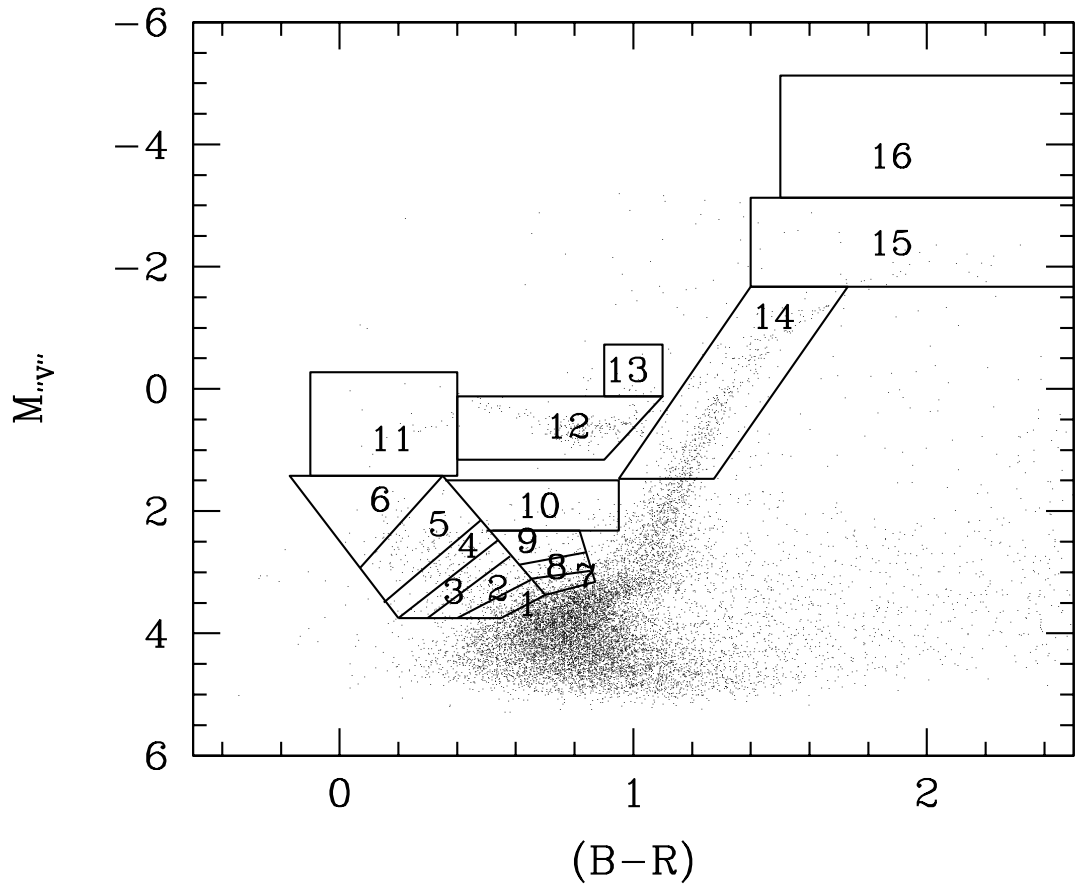,width=16cm}}
\figcaption[cmd_box.eps]{The boxes defined to sample the stellar population
of Draco, overplotted on the CMD. These boxes are used to derive Draco's SFR
with the {\it partial model} method (see text).
\label{cmd_box}}
\end{figure}

Ten partial models have been considered for the analysis of Draco SFH,
covering the age interval from 15 to 2 Gyr. The latter has been assumed the
youngest age present in the CMD because it is the age of the isochrone having
the turn-off point at $''V''\sim 1.5$, corresponding to the upper point of
the Draco blue plume. On the other hand, 16 boxes have been defined in the
CMD, as shown in Fig. \ref{cmd_box}, to characterize the distribution of
stars.  Boxes 1 to 6 sample the younger population, living in the upper
MS. The consequences of this assumption are discussed below. In this analysis
we will assume that all these stars are MS stars and not BS. Boxes 7 to 10
sample the subgiant (SG) region. The information provided by these stars is
complementary to the former one. Fainter stars are not considered because
they are affected by bigger observational effects. In fact, this will limit
the time resolution for the oldest stellar ages. Boxes 11 and 12 sample the
HB. These boxes sample the old stellar population. In particular, if age is
the second parameter, box 11 would contain the oldest pure population. Here,
we will assume that this is the case, but it must be kept in mind that the
nature of the second parameter is not fully confirmed and that the blueward
extension of the HB depends on stellar evolution parameters that are not well
controlled. The color uncertainty produced by this is the reason why we
define boxes 11 and 12 to be quite extended in color.

\placefigure{cmd_box}

Box 13 samples intermediate-age to young stars in the core He-burning
phase. They should have a counterpart in the blue plume and the upper SG
region. Region 14 samples the RGB, populated by old and intermediate-age
stars born in fact in the full, 2--15 Gyr interval considered here, as well
as some AGBs in the same age interval. Although it provides nothing in terms
of age resolution when compared with the information contained in other
boxes, the number of stars in box 14 is a strong tool to normalize the full
SFR: independently of the age distribution, the integrated SFR for the old
and intermediate-age stars must be compatible with the number of stars
populating box 14. Box 15 samples intermediate-age AGBs. It should have a
counterpart in the blue plume and the SG region. Finally, box 16 sample young
AGBs. The fact that it contain no stars indicates that the SFR in the last
several million years has been zero or negligible. Linear combinations of
partial models predicting stars in this region will be automatically
rejected.

We have so far commented qualitatively on the different age intervals sampled
by each box defined in the CMD. However, the solution for the SFR is found in
a global way, considering the number of stars in all the boxes. In practice,
it is not reasonable to look for the best solution (the linear coefficients
best reproducing the distribution of stars in the boxes of the observational
CMD), but to search all the solutions providing a stellar distribution
compatible with the observed one within some interval. Here, we have adopted
as good solutions all those producing a number of stars in each box within
$2\sigma$ from the observed value, $\sigma$ for each box being the maximum
between $(N_j^o)^{(1/2)}$ and 1. The reason for adopting a minimum value of 1
for $\sigma$ is to dump small-number effects on the solution.

\begin{figure}
\centerline{\psfig{figure=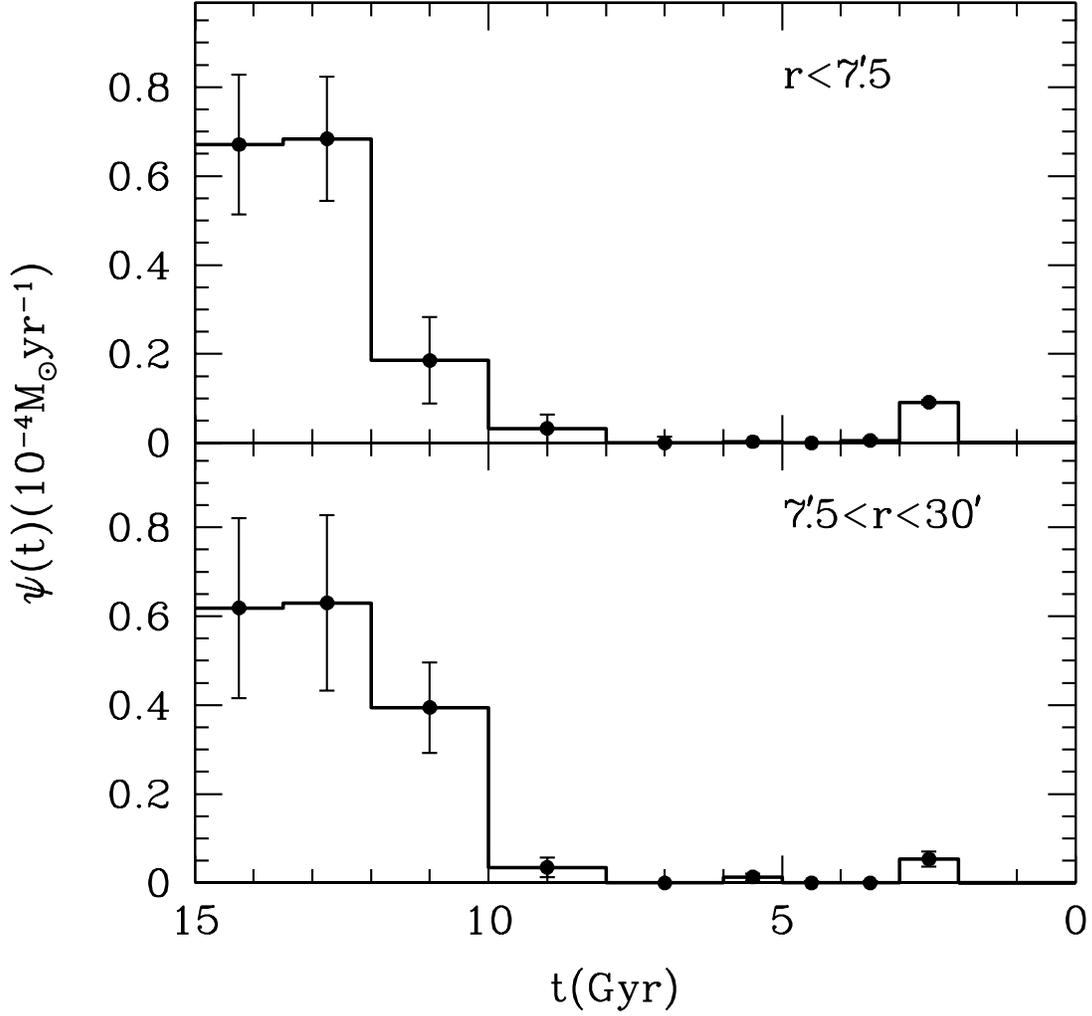,width=16cm}} \figcaption[sfr_pm.eps]{The
solution for the SFR of Draco using the {\it partial model} method. Two
solutions are given. The first for the inner $7\farcm 5$ and the second for
the region between $7\farcm 5$ and $30'$. Error bars account for the
dispersion of possible solutions for each age range.
\label{sfr_pm}}
\end{figure}

\placefigure{sfr_pm}

The resulting SFR is shown in Figure \ref{sfr_pm}. Error bars represent the
sigma value of all the linear combinations producing good solutions for each
age interval. The most relevant fact of the solution is that the dominant
population is old (older than 10 Gyr). It is also remarkable that the SFR
extends down to a few Gyr-old stars, including a jump between 2 and 3
Gyr. The later depends on the assumption we have made that the blue plume is
made up of normal MS stars. If this is wrong, the SFR of intermediate-age to
young stars would be lower than the values obtained. However, the
intermediate-age star formation activity is also supported by the presence of
subgiants and core He-burning stars in the CMD.

This solution cannot be however considered as completely deterministic. In
fact, error bars are in general comparable to the values of $\psi(t)$ for
ages between 3 and 10 Gyr. This, however, does not mean that we know very
little about the solution, but that we cannot determine with a high degree of
confidence whether the most conspicuous population is 12--13.5 or 13.5--15
Gyr old, for example, or whether the jump at 2--3 Gyr is as strong as it
appears in Fig. \ref{sfr_pm} or is more (or less) diluted in time.

Summarizing, the fact that the dominant population in Draco is old may be
considered as a well established result of the present analysis. Also, under
the assumption that the stars populating boxes 1 to 6 (Fig. \ref{cmd_box})
are MS stars rather than blue stragglers, a low rate, intermediate-age
stellar population exist. The jump at 2--3 Gyr is compatible with an
overpopulation of stars in box 4 but mainly comes from the stars populating
box 13. These stars can well be $\sim$ 1.2--1.5 $M_\odot$ in the phase of
core He burning.

The SFR obtained here has been used to calculate the total mass in stars and
stellar remnants ($M_\star$) for the inner $7\farcm 5$ and $30'$ (semi-major
axis). For this, it has been assumed that a fraction 0.8 of the integral of
the SFR remains locked into stars or stellar remnants. The value obtained for
$30'$ can be considered as a good estimate of the total real value, since the
fraction of stars at distances larger than $30'$ is negligible (see
Fig. \ref{densi_p}). Both values are quoted in Table \ref{densi_par} and the
second is used to calculate the dark matter fraction, $\kappa$.

\subsection{The subgiant method}

The turn-off is the most sensitive indicator of stellar age for single
stellar populations. However, in systems such as galaxies, where the stellar
population is composite, the turn-offs of successive stellar generations mix
with each other and with the main sequence of younger stars. In such cases,
the subgiant (SG) region is preferable for measuring stellar ages. It also
has the advantage of being, for a given age and metallicity, above the
corresponding turn-off point, thereby allowing a better signal to noise in
the magnitudes.

Mighell (1997) introduced a simple method based on star counts in the SG
region to determine the SFR. For this purpose, the SG region is divided with
isochrones of a given metallicity and increasing ages. Then the number of
stars are counted between each two isochrones, which provides a present-day
age distribution. Normalized by the stellar lifetime in the SG region as a
function of age, this distribution results in the SFR. In the present case,
we perform this normalization dividing the observed, present-day age
distribution by the distribution corresponding to a synthetic CMD computed
for a constant SFR and the metallicity adopted for Draco.

Resolution is worse for older ages, due both to the increasing photometry
errors and to the decreasing width of the magnitude interval between
isochrones. The method provides a good time resolution up to $\sim 8$ Gyr.

A severe limitation of the method, as it is presently applied, is that it
allows only a constant metallicity for the stellar population. This, however,
is not very important for very low metallicity galaxies such
 as Draco. Also, as a
matter of fact, this method uses only a part of the information that is used
by the partial model method, for which the solution should be
considered less accurate and conclusive. However, in the present case, in
which metallicity evolution is not tested, the subgiant method provides
an alternative approach in which the solution rest only on the SG stars.

In the present analysis, and for the same reasons as given in \S4.1, the
adopted metallicity was $Z=0.0005$, with an He abundance of
$Y=0.25$. With this input and the Kroupa \etal (1993) IMF, a synthetic CMD
has been computed for constant SFR in the age interval 1--15
Gyr. Observational effects have been simulated as explained in Aparicio \&
Gallart (1995). This is important because errors and completeness
significantly change along the SG region. The simulation has been done six
times with different random number seeds. This allows calculating a
dispersion for the synthetic star counts in each age interval.

\begin{figure}
\centerline{\psfig{figure=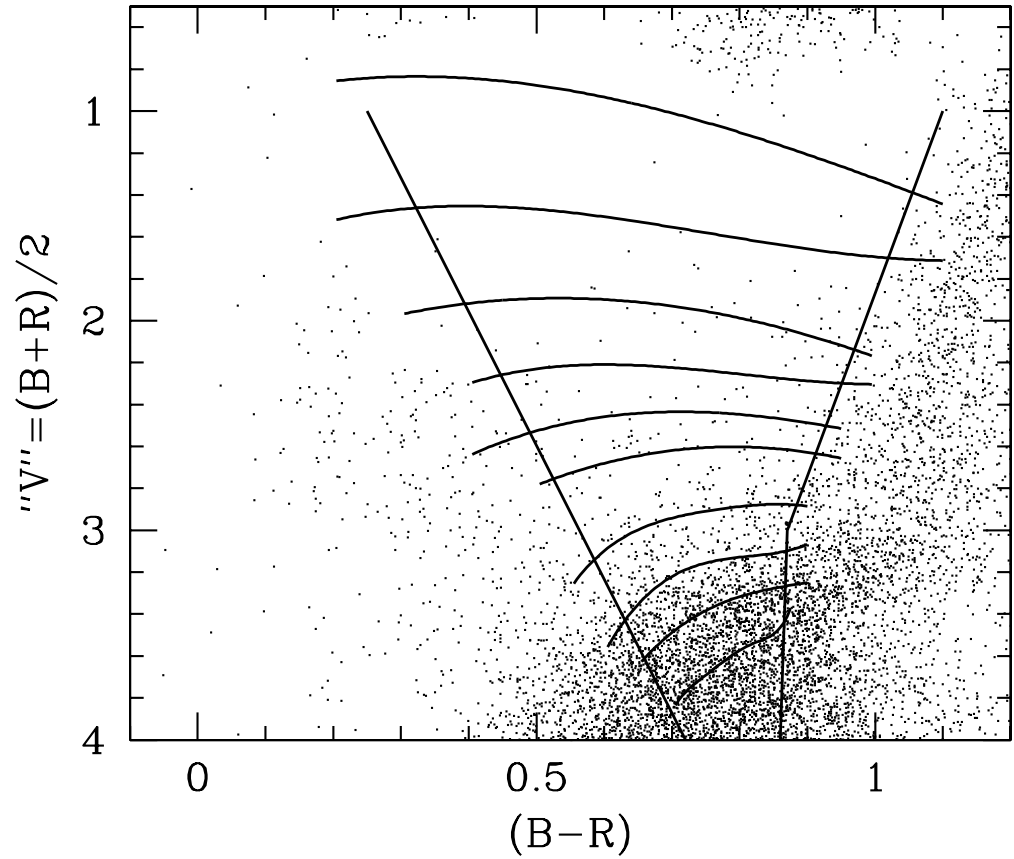,width=16cm}}
\figcaption[cmd_sg.eps]{Subgiant region of Draco's CMD. Isochrones of several
ages are overplotted, corresponding to the ages used to derive the SFR with
the {\it subgiant} method (see text). From up to down, ages are 1.25, 2, 3,
4, 5, 6, 8, 10, 12 and 15 Gyr. Diagonal straight lines show the color
interval used to limit the SG region.
\label{cmd_sg}}
\end{figure}

\begin{figure}
\centerline{\psfig{figure=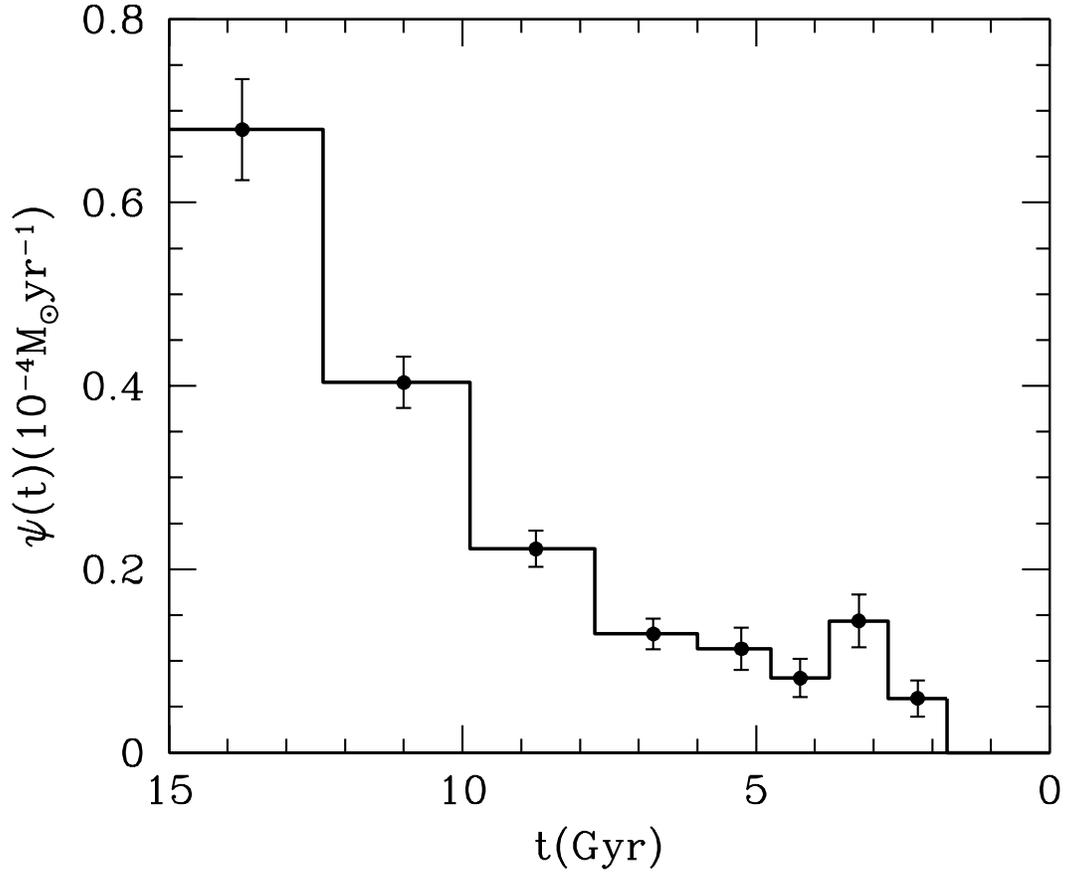,width=16cm}} 
\figcaption[sfr_sg.eps]{The solution for Draco's SFR using the {\it subgiant}
method. The inner $10'$ (semi-major axis) have been considered. 
\label{sfr_sg}}
\end{figure}

\placefigure{cmd_sg}
\placefigure{sfr_sg}

The SG region of Draco's CMD is shown in Figure \ref{cmd_sg}. Isochrones
interpolated from the Padua library (Bertelli \etal 1994) are shown. The
resulting SFR is presented in Fig. \ref{sfr_sg}, which, together with a
dominant old population, shows an intermediate-age population as far as 2 Gyr
ago. To estimate the errors two components have been considered. The first is
just the square root of the star counts for each age interval. It would
account for the fact that we work with a limited number of stars. The second
is the dispersion found for the synthetic star counts and would account for
the random fluctuations in the observational effect simulation.

The most interesting thing is that both the partial model and the 
subgiant methods produce qualitatively similar results, including the young
bump which, although less clear, is also visible in Fig. \ref{sfr_sg},
slightly shifted to older ages. This stress the stability of the solution
and reinforces the capability of both methods to obtain the SFH of nearby
galaxies. The main difference is that the SFR derived by the subgiant
method is more softly decreasing and it maintains a higher value
at intermediate ages. This could be due to observational effects that would
spread the stars in the subgiant region, the overall result being that
fainter stars are moved upwards. In any case, the fact that the  partial
model method uses more information from the CMD eventually makes preferable
the result obtained with it.

\section{Conclusions}

We have presented a photometric and morphological analysis of the
Draco dwarf spheroidal galaxy, as well as a study of its stellar
content and star formation history based on wide field photometry covering
about $1\grados\times1\grados$. 

Two methods ({\it partial model} and {\it subgiant}) have been used to
investigate the SFH of Draco. The most important result is that the stellar
population is fundamentally old (older than $\sim 10$ Gyr). Some 90\% or 75\%
(depending on the method) of the star formation in Draco took place before
$\sim 10$ Gyr ago. After that, the star formation continued at a low rate
and, apparently, a small burst was produced some 2--3 Gyr ago. No star
formation activity is detected in the analysis of the CMD for more recent
epochs.

Several morphological and global parameters have been calculated for
Draco. Its distance is $d=80\pm 7$ kpc and its metallicity, [Fe/H] $=-1.8\pm
0.2$. No metallicity gradient is detected along the galaxy. 

The star surface density distribution of Draco can be fitted by a single
exponential law of scale-length $\alpha=5\farcm 0\pm 0\farcm 1$, for
galacto-centric distances larger than $5'$. Within errors, the same scale is
found for the old (older than $\sim 9$ Gyr) and the young ($\sim 2$--3 Gyr)
stars, indicating either that both populations were formed under the same
kinematic conditions or that any initial difference was afterwards erased.

Evidences have been found of the galaxy extending as far as $42'$ (1 kpc)
from its center. This is in good agreement with the discovery by Piatek \etal
(2001) of stars belonging to Draco beyond the previous tidal radius estimate
by Irwin \& Hatzidimitriou (1995) ($28'$). However, the absence of plateaus
in the density profiles at large galacto-centric distances and the fact that
the scale-length along the semi-major axis and semi-minor axis (re-scaled to
semi-major axis units using the ellipticity of the galaxy) are the same
within errors, indicates that a tidal-tail is not present. More data
are probably required to reach stronger conclusions in this sense.

The central surface magnitude is estimated to be $\mu_{''V''}=24.4\pm
0.5$ and the core radius $r_{\rm c}=7\farcm 5\pm 0\farcm 3$ or
$r_{\rm c}=175\pm 7$ kpc.

The possibility suggested by Kroupa (1997) and Kessel \& Kroupa (1998)
that the high velocity dispersions of dwarf spheroidal galaxies could
be accounted for by adequate spatial orientations rather that being
produced by large amounts of dark matter has been checked. The
conclusion is that this cannot be the case for Draco. An upper limit
to Draco's size along the line of sight is $\sim 14$ kpc; less than
1/3 the value required by Kroupa (1997) models.

\acknowledgements

We are grateful to Dr. Kroupa, who made the results of their tridimensional
models available to us and for several fruitful discussions and to
Dr. G\'omez-Flechoso, for several useful sugestions and discussion on the
dynamical part of the paper.

This article is based on observations made with the 2.5 m Isaac Newton
Telescope operated on the island of La Palma by the ING in the Spanish
Observatorio del Roque de Los Muchachos of the Instituto de Astrof\'\i sica
de Canarias. This research has made use of the NASA/IPAC Extragalactic
Database (NED) which is operated by the Jet Propulsion Laboratory, California
Institute of Technology, under contract with the National Aeronautics and
Space Administration.  This research has been supported by the Instituto de
Astrof\'\i sica de Canarias (grant P3/94), and the General Directorates of
Research of the Kingdom of Spain (grant PI97-1438-C02-01), and of the
autonomous government of the Canary Islands (grant PI1999/008).

\newpage

\begin{deluxetable}{ccccc}
  \tablenum{1}
  \tablewidth{400pt}
  \tablecaption{Journal of observations
  \label{journal}}
  \tablehead{
  \colhead{Date} & \colhead{Draco Field} & \colhead{Time (UT)} &
  \colhead{Filter} &
  \colhead{Exp. time (s)}}
  \startdata
  98.05.29 & A & 00:01 & $B$ & 900 \nl
  98.05.29 & A & 00:19 & $B$ & 900 \nl
  98.05.29 & A & 00:37 & $R$ & 550 \nl
  98.08.20 & A & 21:38 & $R$ & 30 \nl
  98.08.20 & A & 21:42 & $R$ & 900 \nl
  99.06.16 & A & 02:05 & $B$ & 120 \nl
  99.06.16 & A & 02:12 & $R$ & 120 \nl
  00.06.09 & B & 02:29 & $B$ & 900 \nl
  00.06.09 & B & 02:45 & $B$ & 900 \nl	
  00.06.09 & B & 03:01 & $B$ & 900 \nl	
  00.06.09 & B & 03:16 & $B$ & 900 \nl	
  00.06.09 & B & 03:33 & $R$ & 600 \nl	
  00.06.09 & B & 03:33 & $R$ & 600 \nl	
  00.06.09 & B & 03:43 & $R$ & 600 \nl	
  00.06.09 & B & 03:54 & $R$ & 600 \nl	
  00.06.09 & B & 04:05 & $R$ & 600 \nl	
  00.06.09 & B & 04:16 & $R$ & 600 \nl	
  00.06.09 & B & 04:27 & $R$ & 600 \nl	
  00.09.23 & C & 20:24 & $R$ & 120 \nl	
  00.09.23 & C & 20:28 & $R$ & 900 \nl	
  00.09.23 & C & 20:44 & $R$ & 900 \nl	
  00.09.23 & C & 21:00 & $R$ & 900 \nl	
  00.09.23 & C & 21:17 & $B$ & 120 \nl	
  00.09.23 & C & 21:20 & $B$ & 900 \nl	
  00.09.23 & C & 21:36 & $B$ & 900 \nl	
  00.09.23 & C & 21:52 & $B$ & 900 \nl	
  \enddata

  \end{deluxetable}

\newpage

\begin{deluxetable}{ccccc}
\tablenum{2}
\tablewidth{400pt}
\tablecaption{Globular clusters used for the metallicity calibration in the
 $[M_R,(B-R)]$ plane
\label{cum}}
\tablehead{
\colhead{Cluster} & \multicolumn{2}{c}{Metallicity} & \colhead{Distance} &
\colhead{$E(B-V)$} \nl
\colhead{} & \colhead{ZW} & \colhead{CG} & \colhead{} & \colhead{}}
\startdata
NGC 1904  & $-1.7$ & $-1.4$ & $15.59$ & $0.01$ \nl
NGC 4590  & $-2.1$ & $-2.0$ & $15.19$ & $0.05$ \nl
NGC 6723  & $-1.1$ & $-1.0$ & $14.85$ & $0.05$ \nl
\enddata
\end{deluxetable}

\newpage

\begin{deluxetable}{cccc}
\tablenum{3}
\tablewidth{300pt}
\tablecaption{Coefficients of the metallicity scales calibration
\label{coef}}
\tablehead{
\colhead{Metallicity scale} & \colhead{a} & \colhead{b} & 
\colhead{c}}
\startdata
ZW  & $1.924$  & $-5.199$ & \nodata   \nl
CG  & $-3.013$ & $13.192$ & $-15.429$ \nl
\enddata

\end{deluxetable}

\newpage
\clearpage

\begin{deluxetable}{lc}
\tablenum{4}
\tablewidth{300pt}
\tablecaption{Density profile scales
\label{escalas}}
\tablehead{
\colhead{Stellar population} & \colhead{$\alpha (')$}}
\startdata
All stars, full annuli              & $4.7\pm 0.3$  \nl
All stars, semi-major axis          & $5.3\pm 0.5$  \nl
All stars, semi-minor axis$^{(1)}$  & $5.2\pm 0.4$  \nl
MS, full annuli                     & $5.0\pm 0.1$  \nl
MS, semi-major axis                 & $4.8\pm 0.3$  \nl
MS, semi-minor axis$^{(1)}$         & $5.2\pm 0.4$  \nl
RGB, full annuli                    & $5.9\pm 0.4$  \nl
RGB, semi-major axis                & $5.1\pm 0.6$  \nl
RGB, semi-minor axis$^{(1)}$        & $5.9\pm 0.5$  \nl
\enddata
\tablecomments{$^{(1)}$ Parameters given for semi-minor axis have been scaled onto
semi-major axis units; i.e., they have been divided by $b/a$.}
\end{deluxetable}

\newpage
\clearpage

\begin{deluxetable}{lc}
\tablenum{5}
\tablewidth{320pt}
\tablecaption{Morphological and integrated parameters of Draco
\label{densi_par}}
\tablehead{
\colhead{Parameter} & \colhead{Value}}
\startdata
$\alpha_{00}$                            & ${\rm 17^h20^m.3}$       \nl
$\delta_{00}$                            & $57^\circ55'$            \nl
$\mu_{''V'',0}$                          & $24.4\pm 0.5$            \nl  
$\mu_{B,0}$                              & $25.0\pm 0.5$            \nl  
$\mu_{R,0}$                              & $23.9\pm 0.5$            \nl  
$\alpha$                                 & $5\farcm 0\pm 0\farcm 1$ \nl
$r_{\rm c}$                              & $7\farcm 5\pm 0\farcm 3$ \nl
                                         & ($175\pm 7$ kpc)         \nl
$d$                                      & $80\pm 7$ kpc            \nl
[Fe/H]                                   & $-1.8\pm 0.2$            \nl
$''V''_T$                                & $10.1\pm 0.5$             \nl
$B_T$                                    & $10.6\pm 0.5$            \nl
$R_T$                                    & $9.6\pm 0.5$             \nl
$M_{''V''}$                              & $-9.4\pm 0.5$            \nl
$M_{B}$                                  & $-8.9\pm 0.5$            \nl
$M_{R}$                                  & $-9.9\pm 0.5$           \nl
$L_{``V''}$ ($L_\odot$)                  & $4.7\times 10^5$         \nl
$M_{VT}$ ($M_\odot$)                     & $22\times 10^6$          \nl
$M_{VT}/L_{``V''}$ ($M_\odot$/$L_\odot$) & $47$                    \nl
$M_\star(a<7\farcm 5)$ ($M_\odot$)       & $2.1\times 10^5$         \nl
$M_\star(a<30')$ ($M_\odot$)             & $4.3\times 10^5$         \nl
$M_\star(a<42')$ ($M_\odot$)             & $5.3\times 10^5$         \nl
$\kappa=1-(M_\star(a<42')/M_{VT})$       & $0.98$                   \nl
\enddata

\end{deluxetable}

\newpage

\begin{deluxetable}{lcc}
\tablenum{6}
\tablewidth{400pt}
\tablecaption{Comparison of morphological parameters
\label{par_com}}
\tablehead{
\colhead{Reference} & \colhead{$r_{\rm c}$ ($'$)} & \colhead{$r_{\rm t}$ ($'$)}}
\startdata
Hodge (1964)                   & $6.5$        & $26\pm2$     \nl
Irwin \& Hatzidimitriou (1990) & $9.0\pm 0.7$ & $28.3\pm2.4$ \nl   
Lake (1990)                    & $7.0\pm 2.5$ & $55$         \nl
Pryor \& Kormendy (1990)       & $7.1$        & $36$         \nl 
This work                      & $7.5\pm 0.3$ & $42$         \nl
\enddata

\end{deluxetable}

\newpage

\newpage

\newpage

\newpage

\begin{figure}
\centerline{\psfig{figure=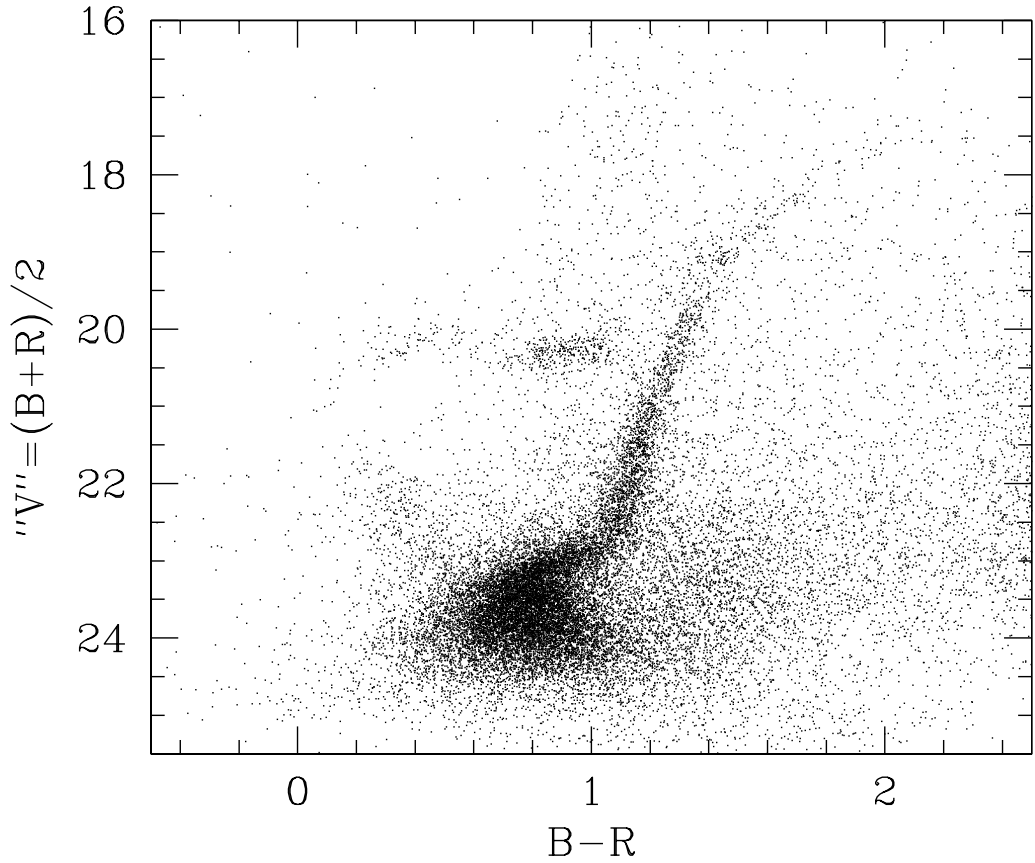,width=16cm}}
\figcaption[cmd.eps]{The CMD of Draco. Stars within the inner $30'$
(semi-major axis) are plotted. Beside a well populated MS and a steep
RGB, indicative of relatively low metallicity, the most significant
features are the extended blue plume and the mainly red HB. Some blue HB
stars and an AGB sequence are also visible.
\label{cmd}}
\end{figure}

\newpage

\newpage

\newpage

\newpage

\newpage
\clearpage

\newpage

\newpage

\newpage

\newpage

\newpage

\newpage

\newpage

\newpage

\end{document}